\documentclass{raa}
\usepackage{graphicx,times}
\usepackage{natbib}
\usepackage{amssymb,amsmath}
\bibpunct{(}{)}{;}{a}{}{,}

\begin{document}

\title{Order and chaos in a galactic model with a strong nuclear bar}

\volnopage{ {\bf 2012} Vol.\ {\bf 12} No. {\bf 5}, 500--512}
   \setcounter{page}{500}

\author{Euaggelos E. Zotos
      }

\institute{Department of Physics, Section of Astrophysics, Astronomy and Mechanics, Aristotle University of Thessaloniki
541 24, Thessaloniki, Greece; {\it evzotos@astro.auth.gr}\\
\vs\no
 {\small Received~~2011 September 20; accepted~~2012~~February 1}}

\abstract{  We use a composite gravitational galactic model
consisting of a disk, a halo, a massive nucleus and a strong
nuclear bar, in order to study the connections between global and
local parameters in a realistic dynamical system. The local model
is constructed from a two-dimensional perturbed harmonic
oscillator and can be derived by expanding the global model in the
vicinity of the central stable Lagrange equilibrium point. The
frequencies of oscillations are not arbitrary, but they are
connected with all the parameters {involved with} the global
model. Moreover, the value of the local energy is also connected
with the value of the global energy. Low and high energy stars in
the global model display chaotic motion. Comparison with previous
research reveals that the presence of the massive nucleus is
responsible for the chaotic motion of the low energy stars. In the
local motion, the low energy stars show interesting resonance
phenomena, {but} the chaotic motion, if any, is negligible. On the
contrary, the high energy stars do not show bounded motion in the
local model. This is an indication of particular activity near the
center of galaxies possessing massive nuclei.  }

\authorrunning{Euaggelos
 E. Zotos}
\titlerunning{Order and Chaos in a Galactic Model with a Strong Nuclear Bar}

\maketitle

\section{Introduction}
\label{sect:intro}

In an earlier work (\citealt{Caranicolas+2002} - hereafter called
Paper P1), we studied the connections between the global and the
local parameters in a barred galactic model. In the present
research, we add a potential of a spherically symmetric nucleus and
thus the total potential becomes
\begin{eqnarray}
\Phi(r,\phi) &=& -\frac{M_{\rm d}}
{\sqrt{r^2 + \alpha^2}} - \frac{M_{\rm b}}{\sqrt{r^2
\left[1 + \left(b^2-1 \right) \sin^2 \phi \right] + c_{\rm b}^2}} \nonumber \\
&&+ \frac{\upsilon_0^2}{2} \ln \left[r^2\left[1 +
\left(\beta^2-1\right) \sin^2 \phi \right] + c_{\rm h}^2\right] -
\frac{M_{\rm n}}{\sqrt{r^2 + c_{\rm n}^2}}\, , \label{eq1}
\end{eqnarray}
where $\left(r,\phi\right)$ are the usual polar coordinates.
Equation~(\ref{eq1}) describes the motion of stars in a barred
galaxy with a disk, a halo, a massive nucleus and a strong nuclear
bar (see below). Here $M_{\rm d}, M_{\rm b}$ and $M_{\rm n}$ are the
masses of the disk, the bar and the nucleus respectively, while
$\alpha, c_{\rm b}, c_{\rm h}$ and $c_{\rm n}$ represent the scale
lengths of the disk, the bar, the halo and the nucleus respectively.
The strength of the nuclear bar is represented by the parameter $b$
$(b>1)$, while the flattening parameter of the halo is represented
by the parameter $\beta$. Moreover, the parameter $\upsilon_0$ is
used for the consistency of the galactic units.

In the system of galactic units used in this article, the unit of
length is 1~kpc, the unit of time is 0.97746 $\times$ $10^7$~yr and
the unit of mass is 2.325 $\times$ $10^7$~$M_{\odot}$. The velocity
and the angular velocity units are 10~km~s$^{-1}$ and
10~km~s$^{-1}$~kpc$^{-1}$ respectively, while $G$ is equal to unity.
Our test particle is a star of unit mass $(m=1)$. Therefore, the
energy unit (per unit mass) is 100~km$^2$~s$^{-2}$. In these units
the values of the parameters{ involved} are: $\alpha=8, \beta=1.3,
b=2, \upsilon_0=15, M_{\rm d}=9500, M_{\rm b}=3000, M_{\rm n}=400,
c_{\rm b}=1.5, c_{\rm n}=0.25$ and $c_{\rm h}=8.5$. The values of
the above dynamical parameters remain constant during this research.

We shall consider the case when the bar rotates clockwise, at a
constant angular velocity $\Omega_{\rm b}$. The corresponding
Hamiltonian, which is the well known Jacobi integral, in the
rectangular Cartesian coordinates $\left(x,y\right)$ is
\begin{eqnarray}
H_{\rm J} &=& \frac{1}{2}\left(p_x^2 + p_y^2\right) + \Phi(x,y) - \frac{1}{2}\Omega_{\rm b}^2\left(x^2 + y^2\right) \nonumber \\
&=& \frac{1}{2}\left(p_x^2 + p_y^2\right) + \Phi_{\rm eff}(x,y) \nonumber \\
&=& E_{\rm J},
\label{eq2}
\end{eqnarray}
where $p_x$ and $p_y$ are the momenta per unit mass, conjugate to
$x$ and $y$ respectively and
\begin{eqnarray}
\Phi_{\rm eff}(x,y) &=&
-\frac{M_{\rm d}}{\sqrt{x^2
+ y^2 + \alpha^2}} - \frac{M_{\rm b}}{\sqrt{x^2 + b^2 y^2 + c_{\rm b}^2}} +
\frac{\upsilon_0^2}{2} \ln\left[x^2 + \beta y^2 + c_{\rm h}^2\right] \nonumber \\
&&- \frac{M_{\rm n}}{\sqrt{x^2 + y^2 + c_{\rm n}^2}} -
\frac{1}{2}\Omega_{\rm b}^2\left(x^2 + y^2\right), \label{eq3}
\end{eqnarray}
is the effective potential, while $E_{\rm J}$ is the numerical value
of the Jacobi integral. If we expand the effective
potential~(\ref{eq3}) in a Taylor series near the center, we shall
obtain a potential describing local motion.

The motivation of the present work is twofold: (i) to investigate
the properties of global and local motion in the corresponding
potentials. In particular, we shall express the coefficients of the
local potential in terms of the global physical quantities entering
the potential~(\ref{eq3}). A connection between the values of the
global and the local energies will also be presented. (ii) to
compare our numerical results with those obtained in Paper P1, where
we only{ had} the nuclear bar, while the massive nucleus was absent.

The present paper is organized as follows: In Section~2 we study the
properties of motion in the global model. The local potential, the
connection between the local and the global parameters and the
properties of the local motion, are presented in Section~3. We close
with a discussion and the conclusions of this research, which are
given in Section~4.

\section{Properties of motion in the global model}
\label{sect:prop}

\begin{figure}

\vs \centering
\resizebox{0.5\hsize}{!}{\rotatebox{0}{\includegraphics*{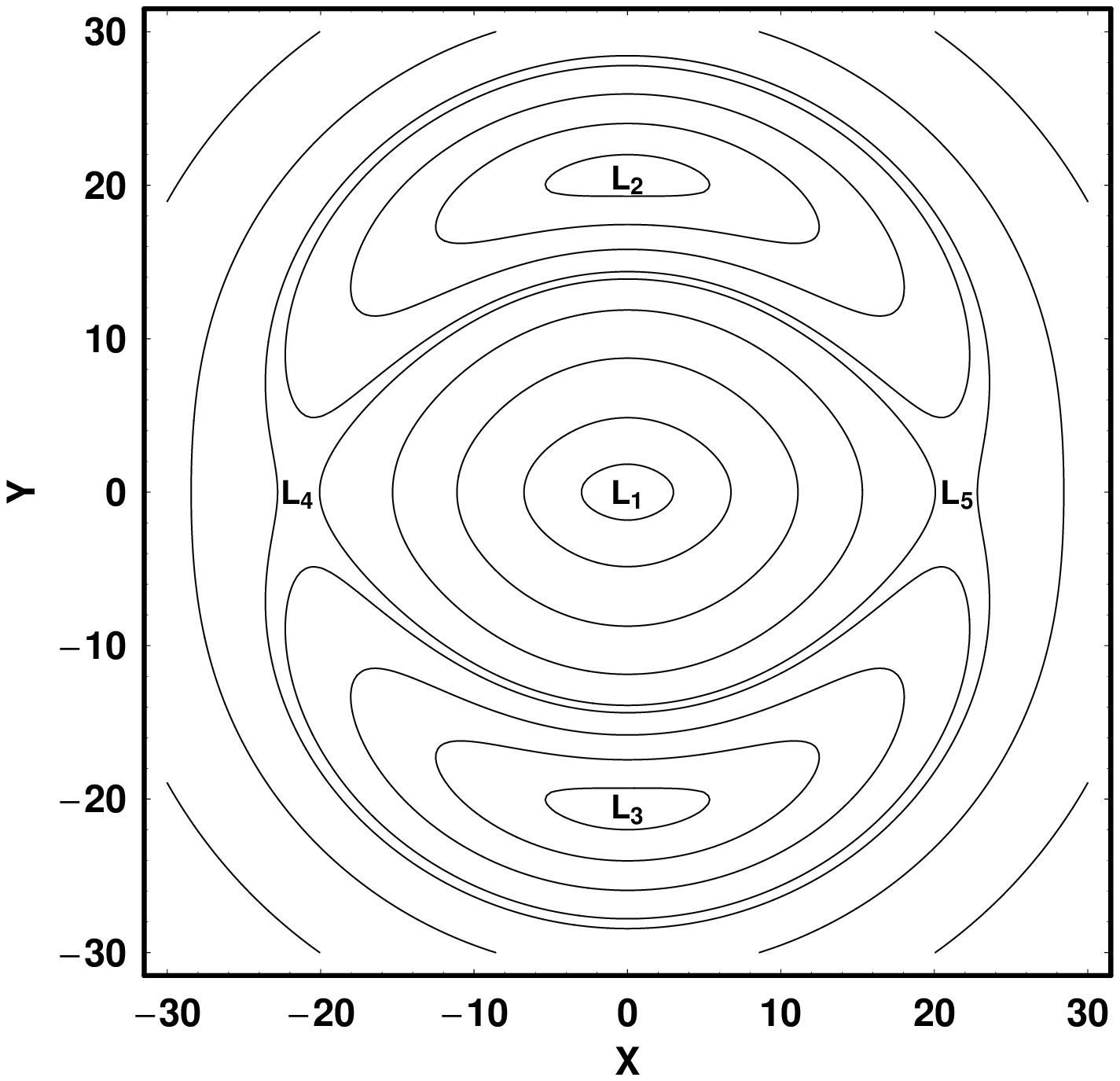}}}

\vspace{-3mm} \caption{\baselineskip 3.6mm Contours of the constant
effective potential~(\ref{eq3}). The values of all the parameters
are given in the text. The five Lagrange points are labeled $L_1$ to
$L_5$.\label{fig1}}
\end{figure}

In Figure~\ref{fig1} we can see the contours of the constant
effective potential~(\ref{eq3}). The value of $\Omega_{\rm b}$ is
1.25 in the above mentioned galactic units. This value corresponds
to 12.5\,km~s$^{-1}$~kpc$^{-1}$. One observes that there are five
stationary points, labeled $L_1$ to $L_5$, at which
\begin{equation}
\frac{\partial \Phi_{\rm eff}}{\partial x} = 0\, , \qquad
\frac{\partial \Phi_{\rm eff}}{\partial y} = 0\, .  \label{eq4}
\end{equation}
These points are called Lagrange points. The central stationary
point $L_1$ is the minimum of $\Phi_{\rm eff}$. At the other four
points $L_2, L_3, L_4$ and $L_5${,} it is possible for the test
particle to travel in a circular orbit while appearing to be
stationary in the rotating frame. For this orbit, the centrifugal
and the gravitational force precisely balance. The stationary points
$L_4$ and $L_5$ on the $x$ axis are saddle points, while $L_2$ and
$L_3$ are the maxima of the effective potential. The annulus bounded
by the circles through $L_2, L_3$ and $L_4, L_5$ is known as the
``region of corotation" (see \citealt{Binney+Tremaine+2008}). It is
also important to note that the region of corotation in our
dynamical system is located somewhere {in} the outer parts of the
galaxy.

We shall now proceed to study the properties of motion in the
potential~(\ref{eq3}). We solve the equations of motion in the
$(x,y)$ plane for the potential~(\ref{eq3}) in which the nuclear bar
rotates independently around the $z$ axis. Assuming a clockwise
rotation with a pattern angular velocity $\Omega_{\rm b}$, one can
write the equations of motion in the form
\begin{equation}
\ddot{r} = - \nabla \Phi_{\rm eff} - 2\left(\Omega_{\rm b} \times
\dot{r} \right) + |\Omega_{\rm b}|^2 r \, .
\end{equation}
Decomposing into its $x$ and $y$ parts, we obtain
\begin{eqnarray}
\ddot{x} &=& - \frac{\partial \Phi_{\rm eff}}{\partial x}
 - 2\Omega_{\rm b} \dot{y} + \Omega_{\rm b}^2 x \, , \nonumber \\
\ddot{y} &=& - \frac{\partial \Phi_{\rm eff}}{\partial y}
 + 2\Omega_{\rm b} \dot{x} + \Omega_{\rm b}^2 y\, ,
\label{eq6}
\end{eqnarray}
where the dot indicates derivative with respect to the time.

All the numerical calculations are based on the numerical
integration of the equations of motion~(\ref{eq6}), which was made
using a Bulirsh-St\"{o}er routine in Fortran 95, with double
precision in all subroutines. The accuracy of the calculations was
checked by the constancy of the Jacobi integral~(\ref{eq2}), which
was conserved up to the eighteenth significant figure.

\begin{figure}\vs
\centering
\resizebox{0.7\hsize}{!}{\rotatebox{270}{\includegraphics*{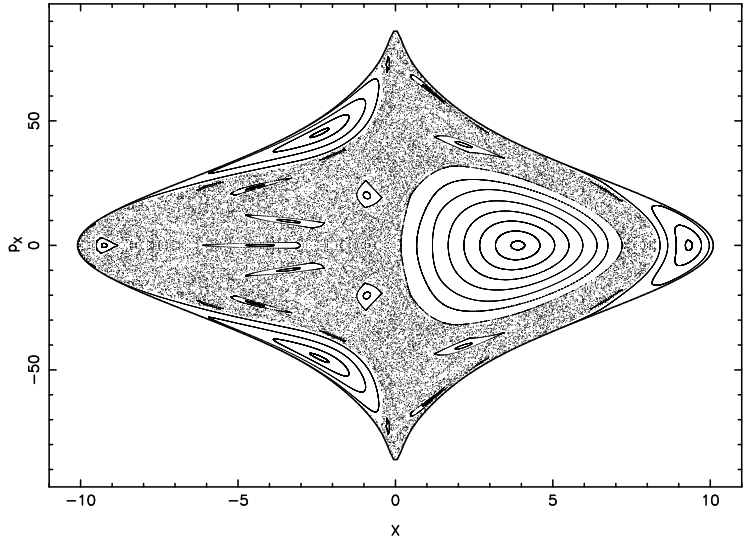}}}

\vspace{-3mm} \caption{\baselineskip 3.6mm The $x-p_x$ Poincar\'{e}
phase plane for the global Hamiltonian~(\ref{eq2}), when $E_{\rm
J}=-570$. The values of all the other parameters are as in
Fig.~\ref{fig1}.\label{fig2}}
\end{figure}

\begin{figure*}

\vs \centering
\resizebox{0.8\hsize}{!}{\rotatebox{0}{\includegraphics*{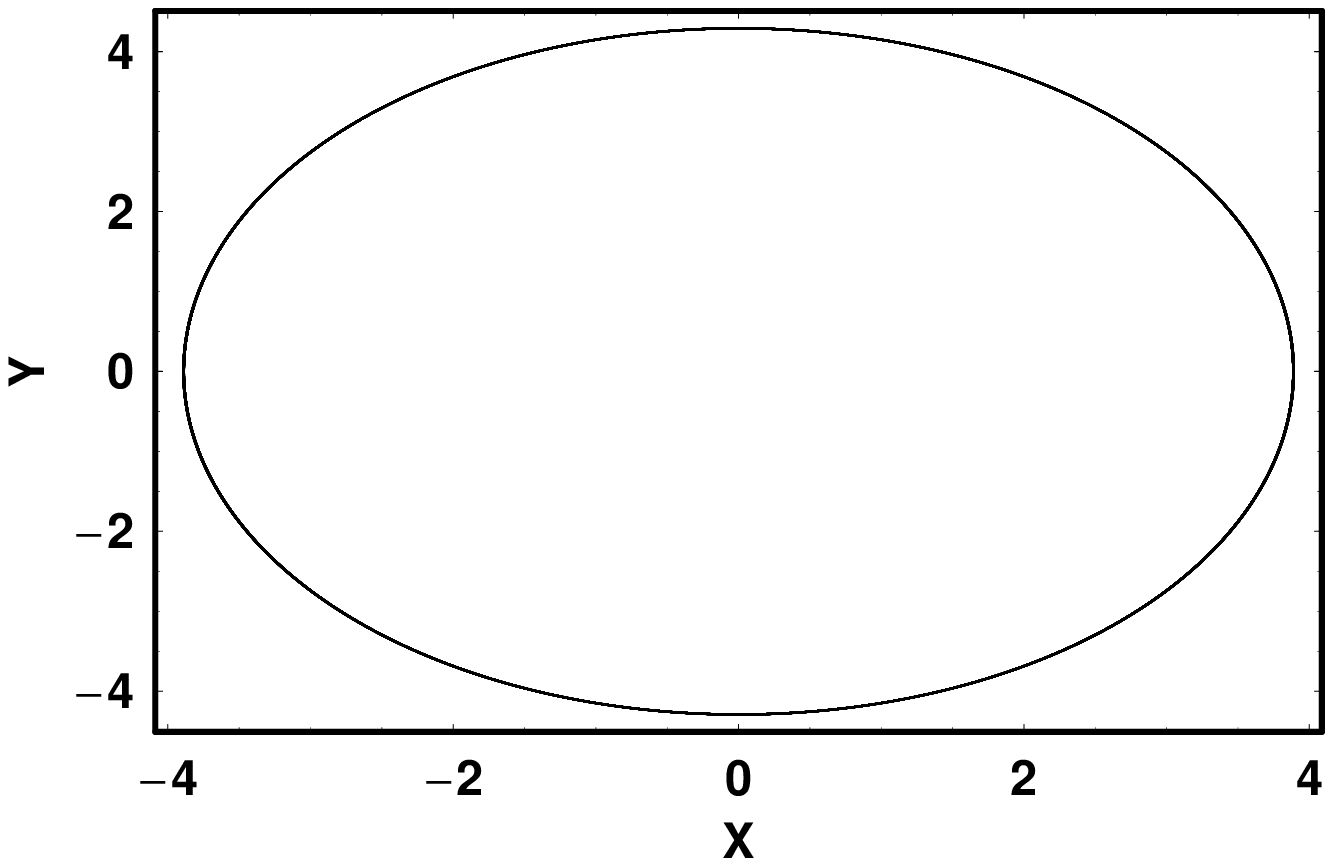}}
                         \rotatebox{0}{\includegraphics*{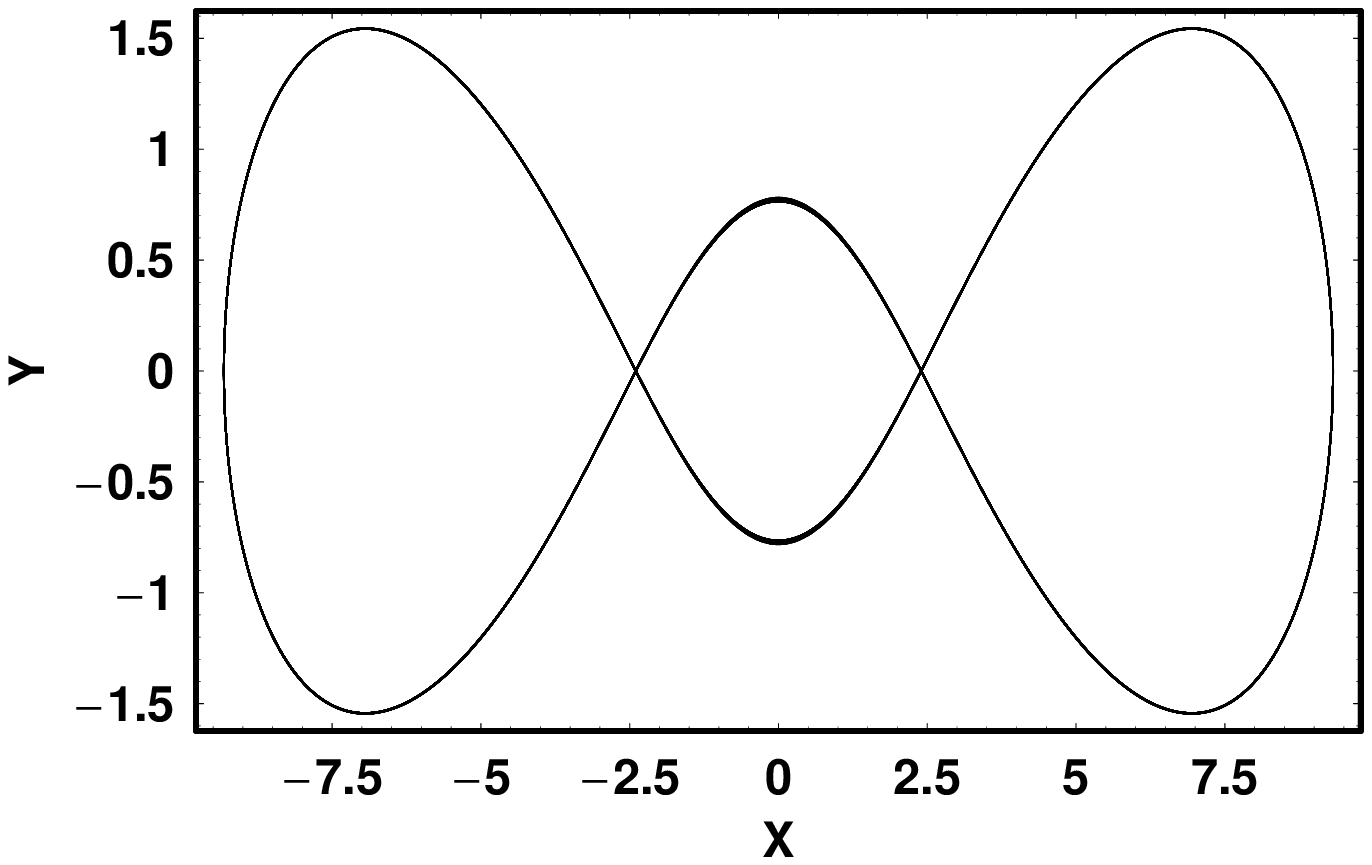}}}

\begin{minipage}{40mm}\centering

\hs {\fns(a)}
\end{minipage}\hspace{30mm}
\begin{minipage}{21mm}
\centering {\fns(b)}~~~~~~~~\end{minipage}\vs

\resizebox{0.8\hsize}{!}{\rotatebox{0}{\includegraphics*{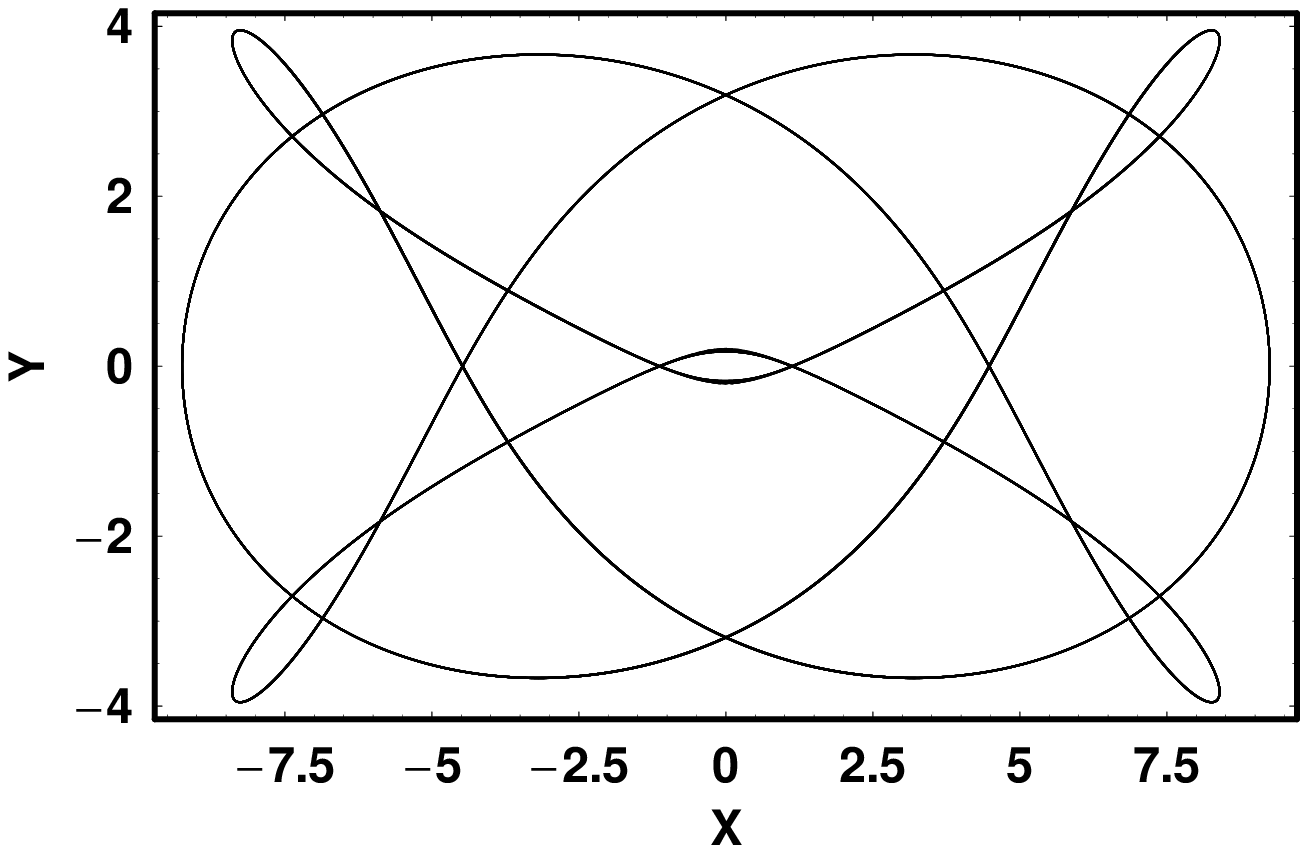}}
                         \rotatebox{0}{\includegraphics*{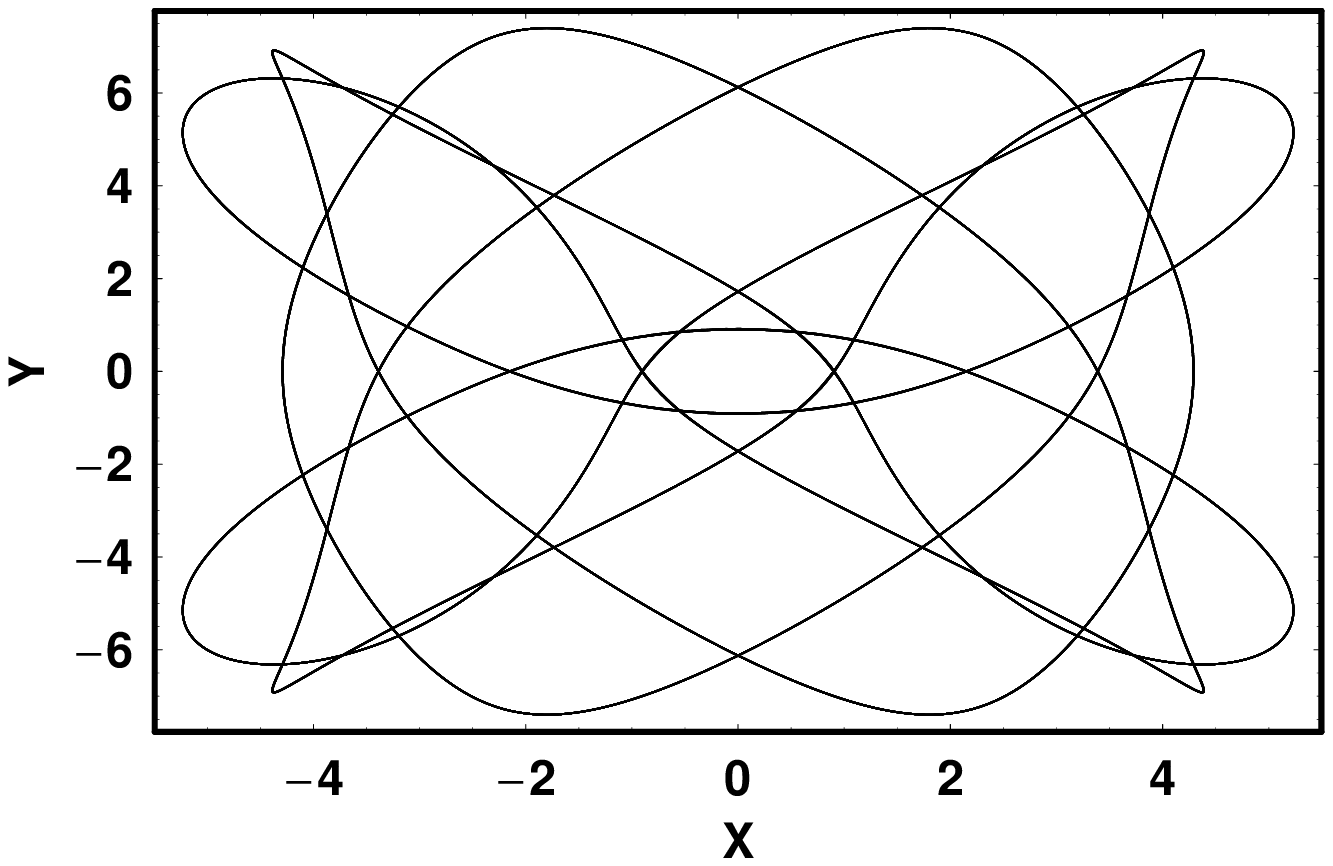}}}

\begin{minipage}{40mm}\centering

\hs {\fns(c)}
\end{minipage}\hspace{30mm}
\begin{minipage}{21mm}
\centering {\fns(d)}~~~~~~~~\end{minipage} \vs

\resizebox{0.8\hsize}{!}{\rotatebox{0}{\includegraphics*{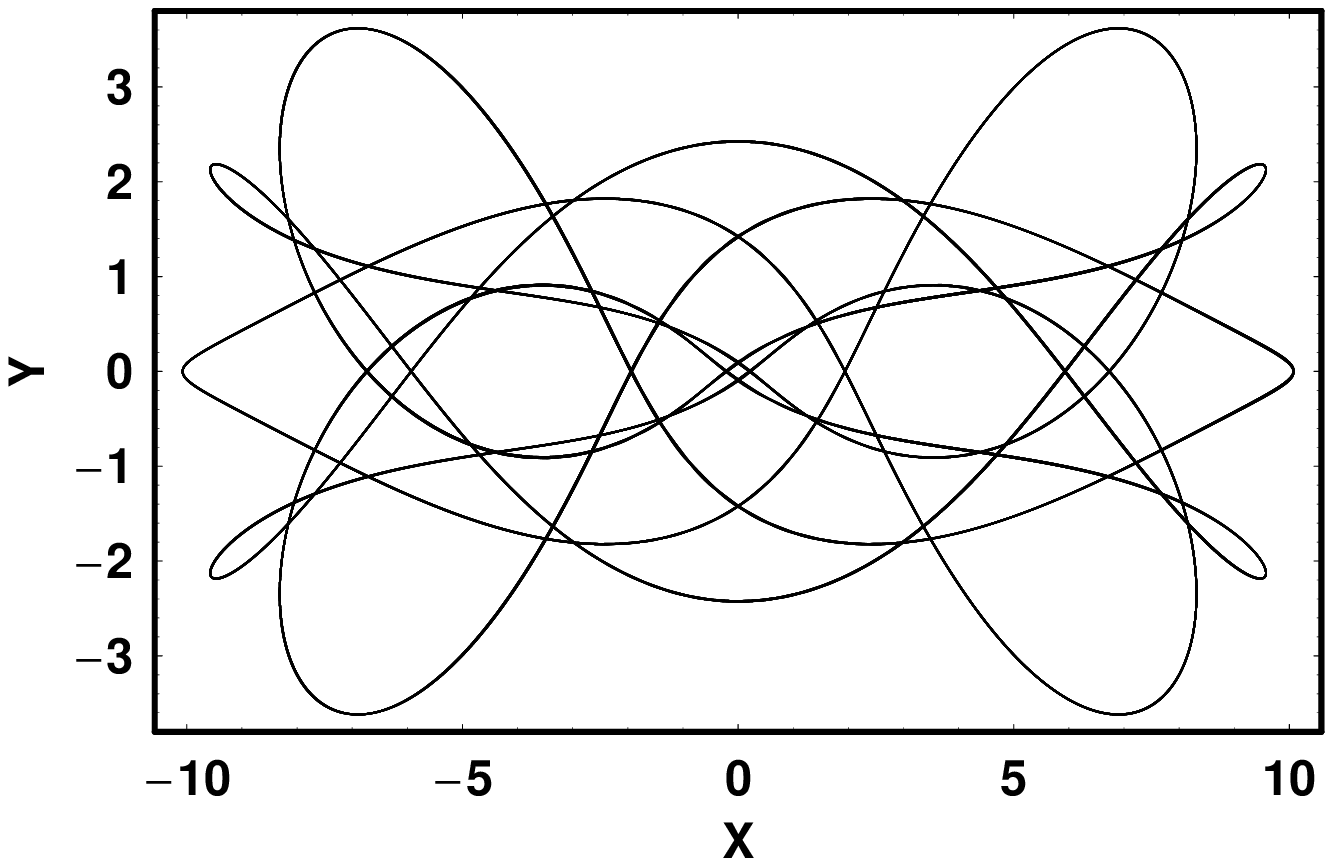}}
                         \rotatebox{0}{\includegraphics*{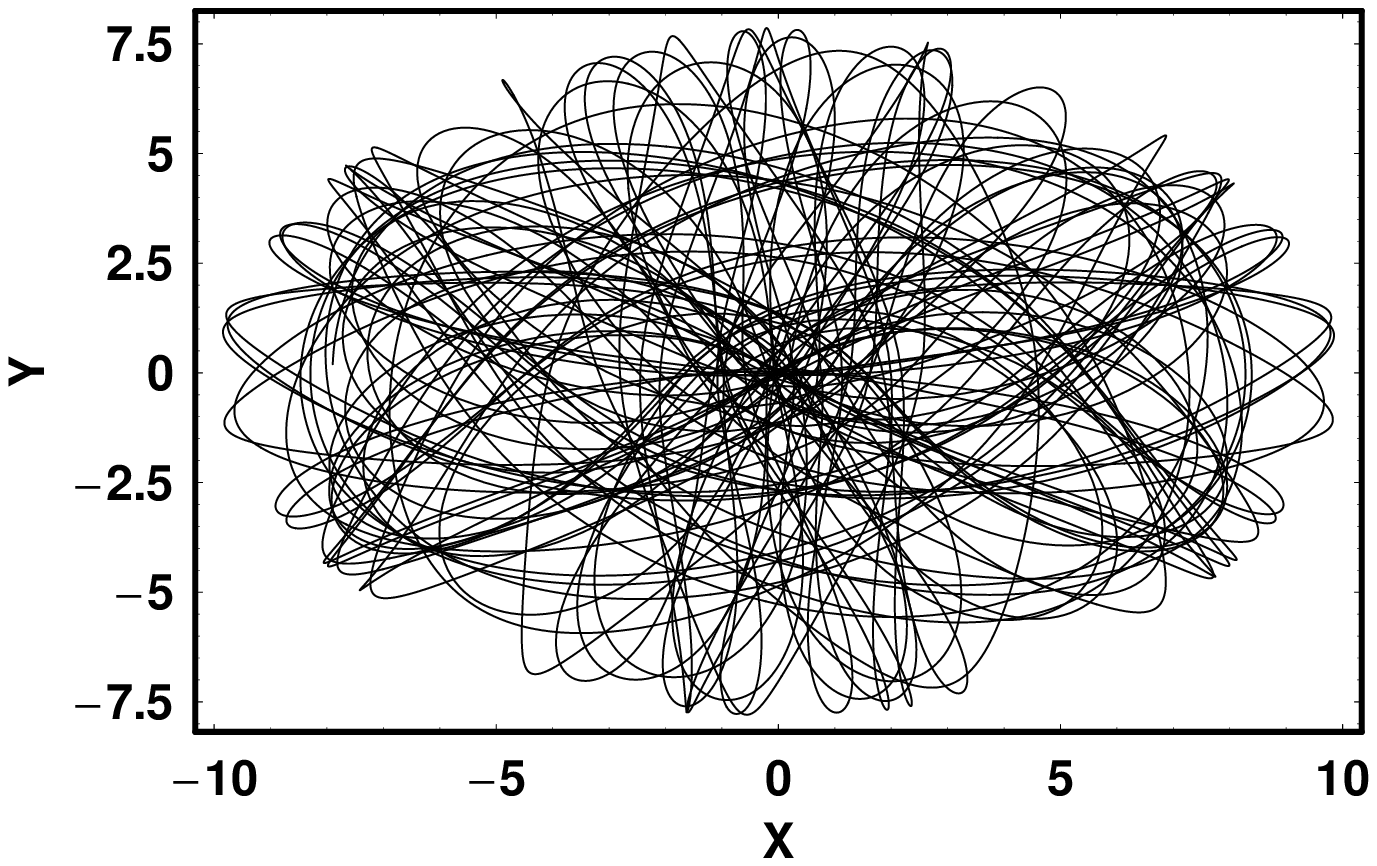}}}

\begin{minipage}{40mm}\centering

\hs {\fns(e)}
\end{minipage}\hspace{30mm}
\begin{minipage}{21mm}
\centering {\fns(f)}~~~~~~~~\end{minipage}

 \caption{\baselineskip 3.6mm (a)--(f): Six typical
orbits in the global Hamiltonian~(\ref{eq2}). The values of all the
other parameters are as in Fig.~\ref{fig2}.\label{fig3}}
\end{figure*}

In order to visualize the properties of motion, we shall use the
classical method of the $x-p_x$, $y=0$, $p_y>0$ Poincar\'{e} phase
plane of the Hamiltonian~(\ref{eq2}). Figure~\ref{fig2} shows the
structure of this phase plane for the global
Hamiltonian~(\ref{eq2}), when $E_{\rm J}=-570$. The particular value
of the energy $E_{\rm J}$ was chosen so that in the phase plane
$x_{\rm max} \simeq 10$. There are regular orbits, forming the
``right" retrograde set of invariant curves, as well as a triple set
of islands. Furthermore, one can identify several sets of smaller
secondary invariant curves embedded {in} the chaotic sea, which
represent resonant orbits of higher multiplicity. In addition to the
regular region, there is a large unified chaotic domain formed by
the chaotic orbits of the dynamical system. It is
important to point out here that this chaotic area is obviously
larger than that observed in the case in Paper P1, where the massive
nucleus was absent. Also note that the presence of the massive
nucleus gives rise to a variety of secondary resonances. If we set
$y=p_y=0$ in Equation~(\ref{eq2}), we obtain the limiting curve in
the $x-p_x$ phase plane, which is the curve containing all the
invariant curves for a given value of the Jacobi integral $E_{\rm
J}$. The limiting curve which is the outermost curve shown in
Figure~\ref{fig2} is defined by the equation
\begin{equation}
\frac{1}{2}p_x^2 + \Phi_{\rm eff}(x) = E_{\rm J}\, . \label{eq7}
\end{equation}

Figure~\ref{fig3}(a)--(f) shows six typical orbits of the global
Hamiltonian system. The orbit shown in Figure~\ref{fig3}(a)
produce{s} one of the ``right" invariant curves. This orbit is a 1:1
resonant periodic orbit starting at the stable retrograde periodic
point and it is nearly circular. Such orbits support the disk. The
initial conditions are: $x_0=3.89$, $y_0=0$, $p_{x0}=0$. The orbit
shown in Figure~\ref{fig3}(b) produces the set of the three outer
islands in the phase plane of Figure~\ref{fig2}. This periodic orbit
belongs to the family of the 1:3 resonant orbits. We observe that
the shape of this orbit is elongated and therefore it supports the
nuclear bar. The initial conditions are: $x_0=9.32$, $y_0=0$,
$p_{x0}=0$. The orbit depicted in Figure~\ref{fig3}(c) produces
t{w}o of the three elongated islands of invariant curves which are
embedded in the chaotic sea. This periodic orbit is characteristic
of the 3:5 resonance and it has initial conditions: $x_0=-9.25$,
$y_0=0$, $p_{x0}=0$. In Figure~\ref{fig3}(d) one can see a typical
example of a periodic orbit which belongs to the family of 5:7
resonant orbits. The initial conditions are: $x_0=-4.29$, $y_0=0$,
$p_{x0}=0$. The orbit shown in Figure~\ref{fig3}(e) is a complicated
periodic orbit characteristic of the 5:9 resonance and produces a
set of nine small islands of invariant curves inside the vast
chaotic domain. This orbit has initial conditions: $x_0=-0.23$,
$y_0=0$, $p_{x0}=73$. It is evident that the last four types of
periodic orbits and also the quasi-periodic orbits which belong to
each family support both the disk and the bar structure of the
galaxy. Finally, in Figure~\ref{fig3}(f) we see a chaotic orbit with
initial conditions: $x_0=-7.9$, $y_0=0$, $p_{x0}=0$. The initial
value of $p_{y0}$ was found in every case from the Jacobi
integral~(\ref{eq2}). All orbits shown in Figure~\ref{fig3}(a)--(f)
were calculated for a time period of 100 time units.

\begin{figure}

\vs \centering
\resizebox{0.7\hsize}{!}{\rotatebox{270}{\includegraphics*{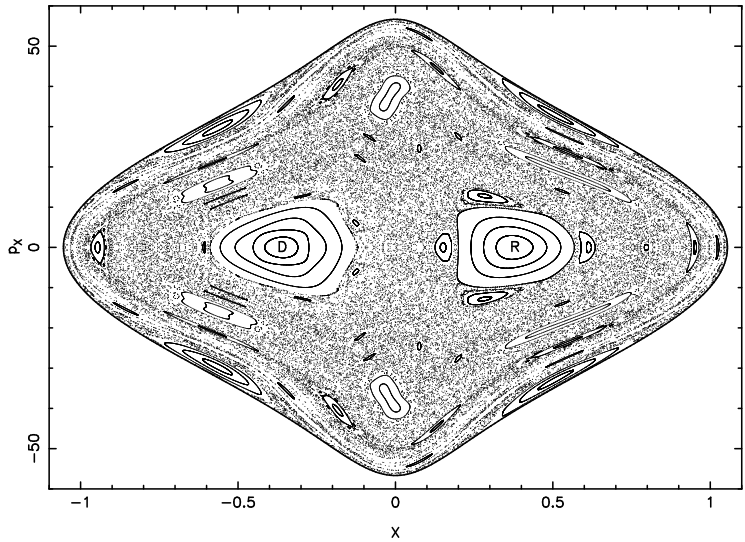}}}

\begin{minipage}{130mm}\caption{\label{fig4}The $x-p_x$ Poincar\'{e} phase plane, when
$E_{\rm J}=-2700$. Details are given in the text.}\end{minipage}
\end{figure}

\begin{figure*}
\centering

\vs
\resizebox{0.8\hsize}{!}{\rotatebox{0}{\includegraphics*{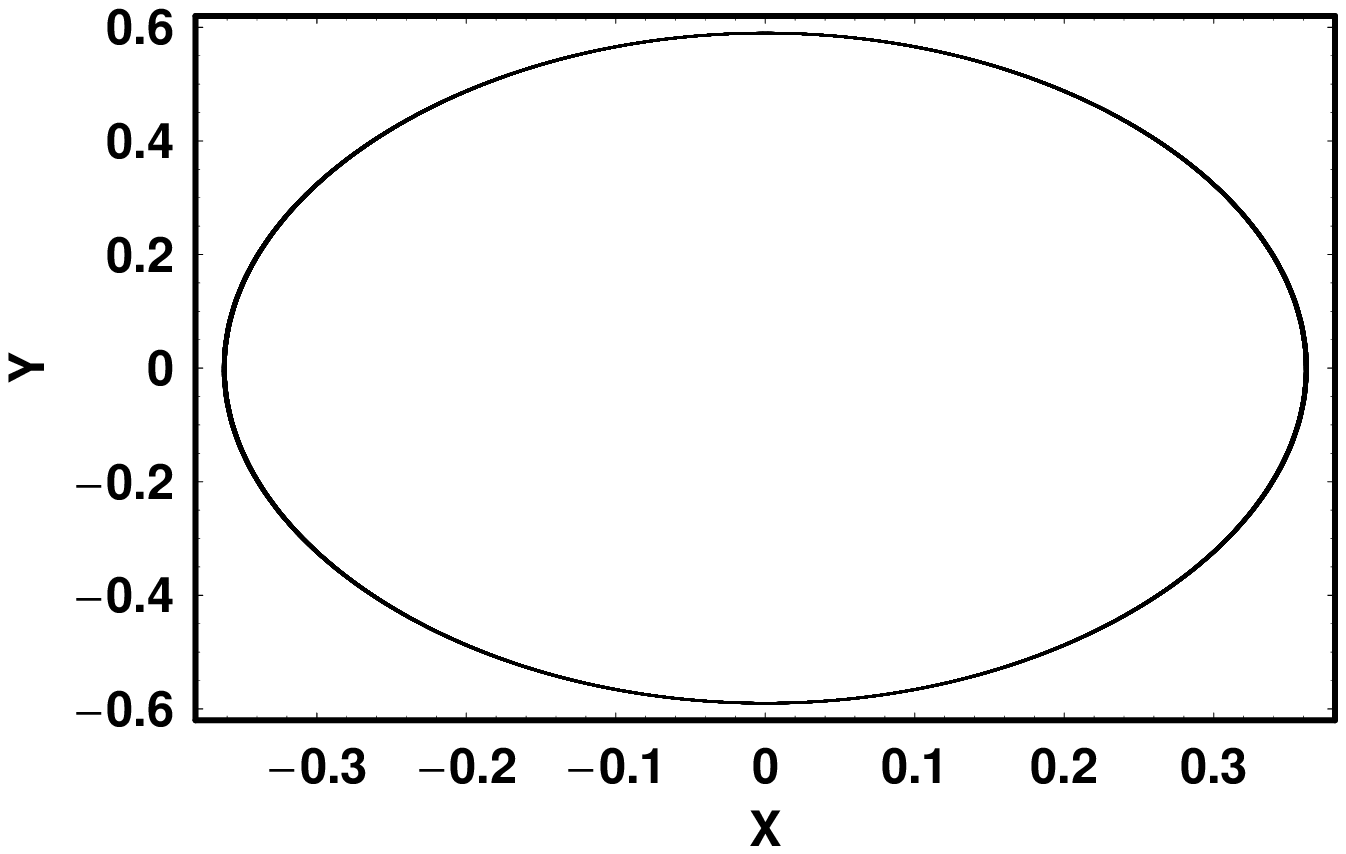}}
                         \rotatebox{0}{\includegraphics*{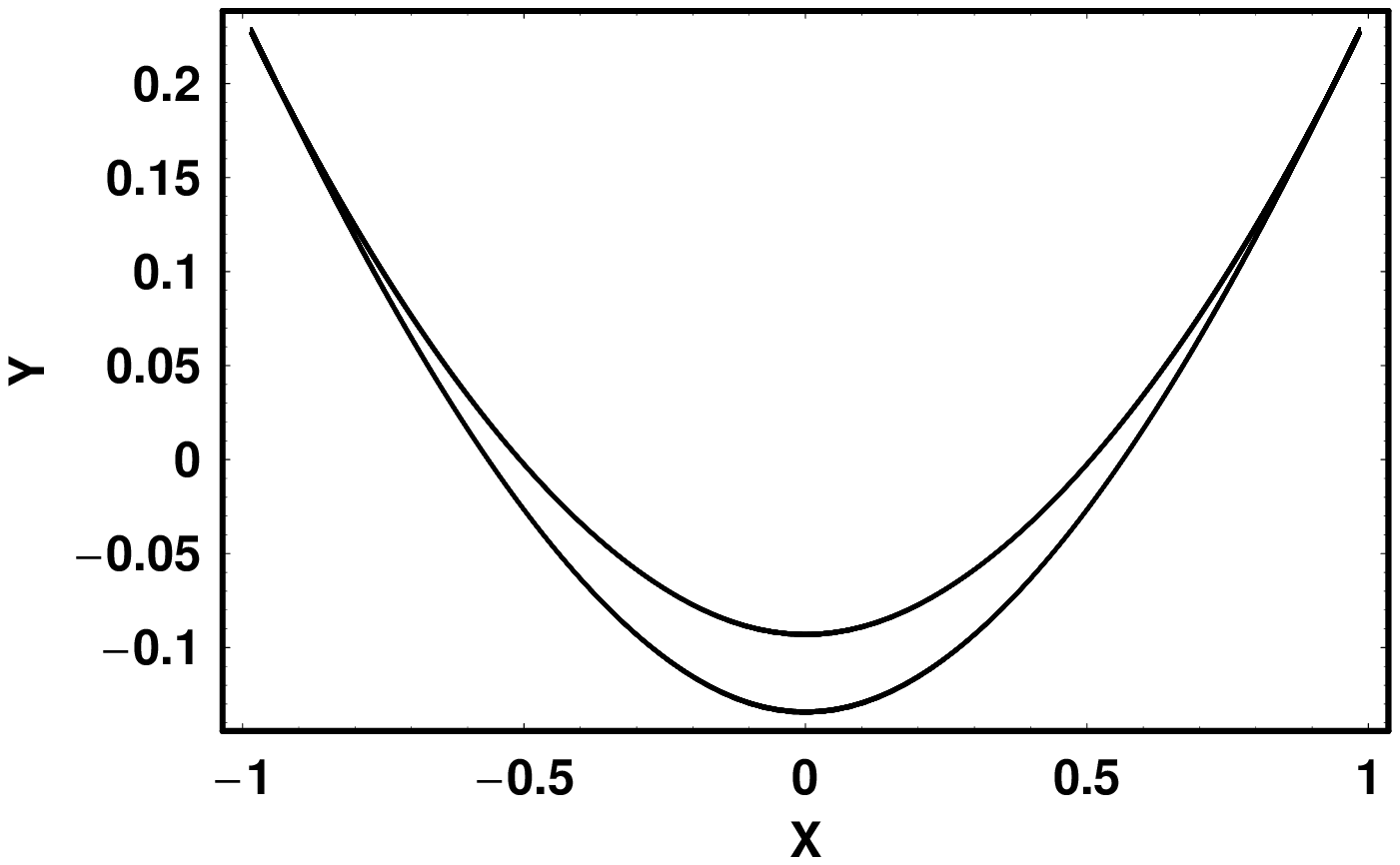}}}

\begin{minipage}{40mm}\centering

\hs {\fns(a)}
\end{minipage}\hspace{30mm}
\begin{minipage}{21mm}
\centering {\fns(b)}~~~~~~~~\end{minipage}

\resizebox{0.8\hsize}{!}{\rotatebox{0}{\includegraphics*{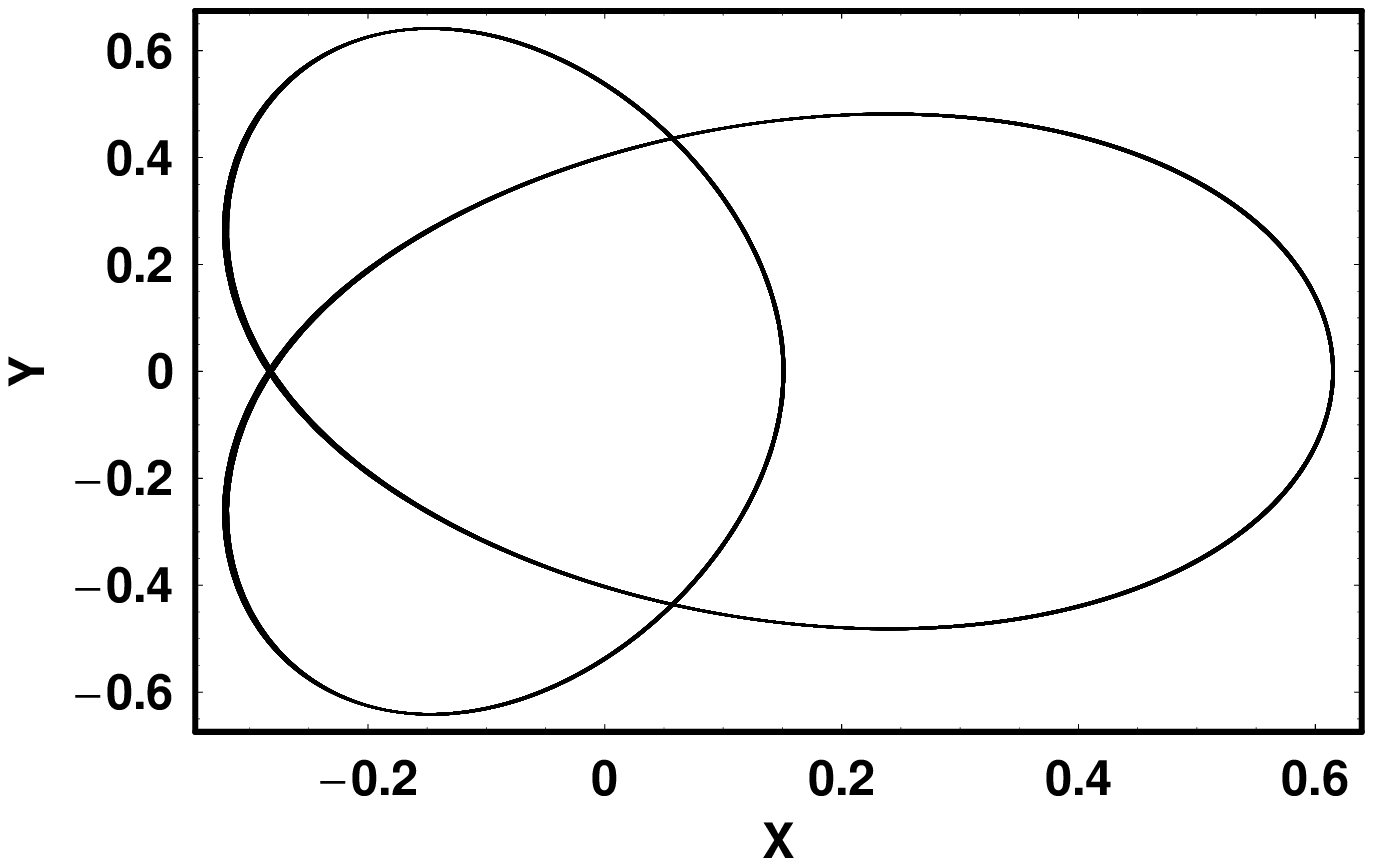}}
                         \rotatebox{0}{\includegraphics*{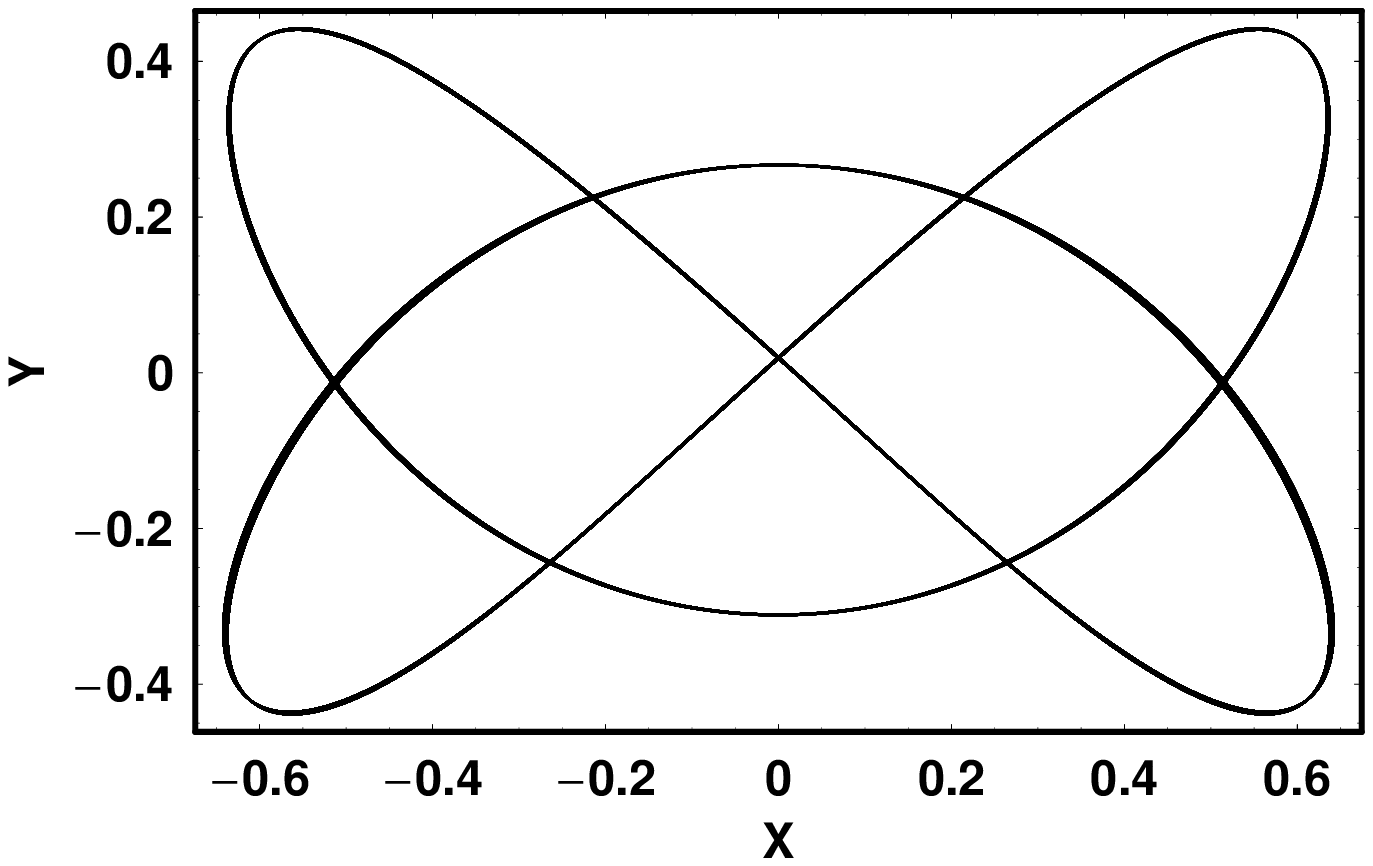}}}

\begin{minipage}{40mm}\centering

\hs {\fns(c)}
\end{minipage}\hspace{30mm}
\begin{minipage}{21mm}
\centering {\fns(d)}~~~~~~~~\end{minipage}

\resizebox{0.8\hsize}{!}{\rotatebox{0}{\includegraphics*{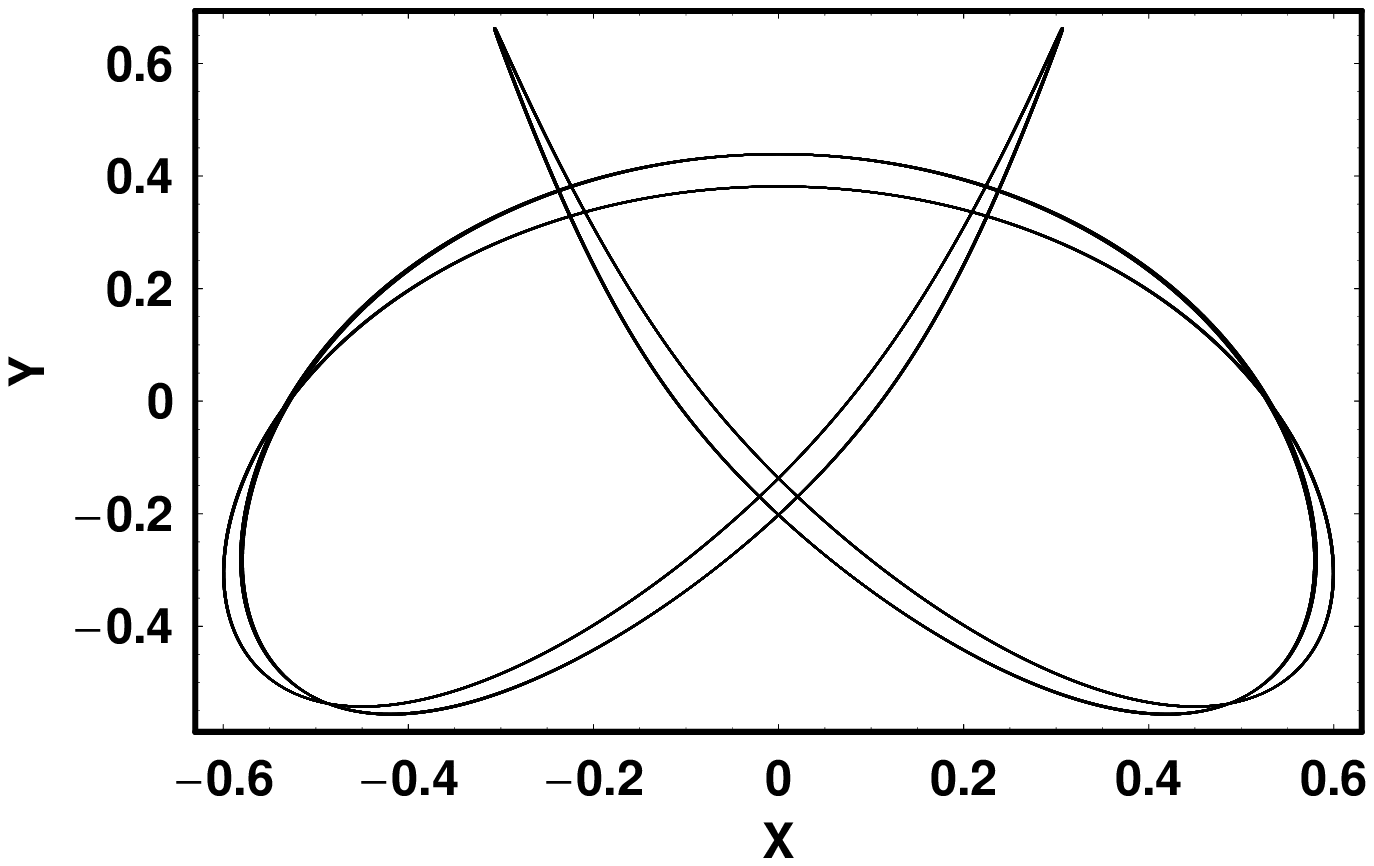}}
                         \rotatebox{0}{\includegraphics*{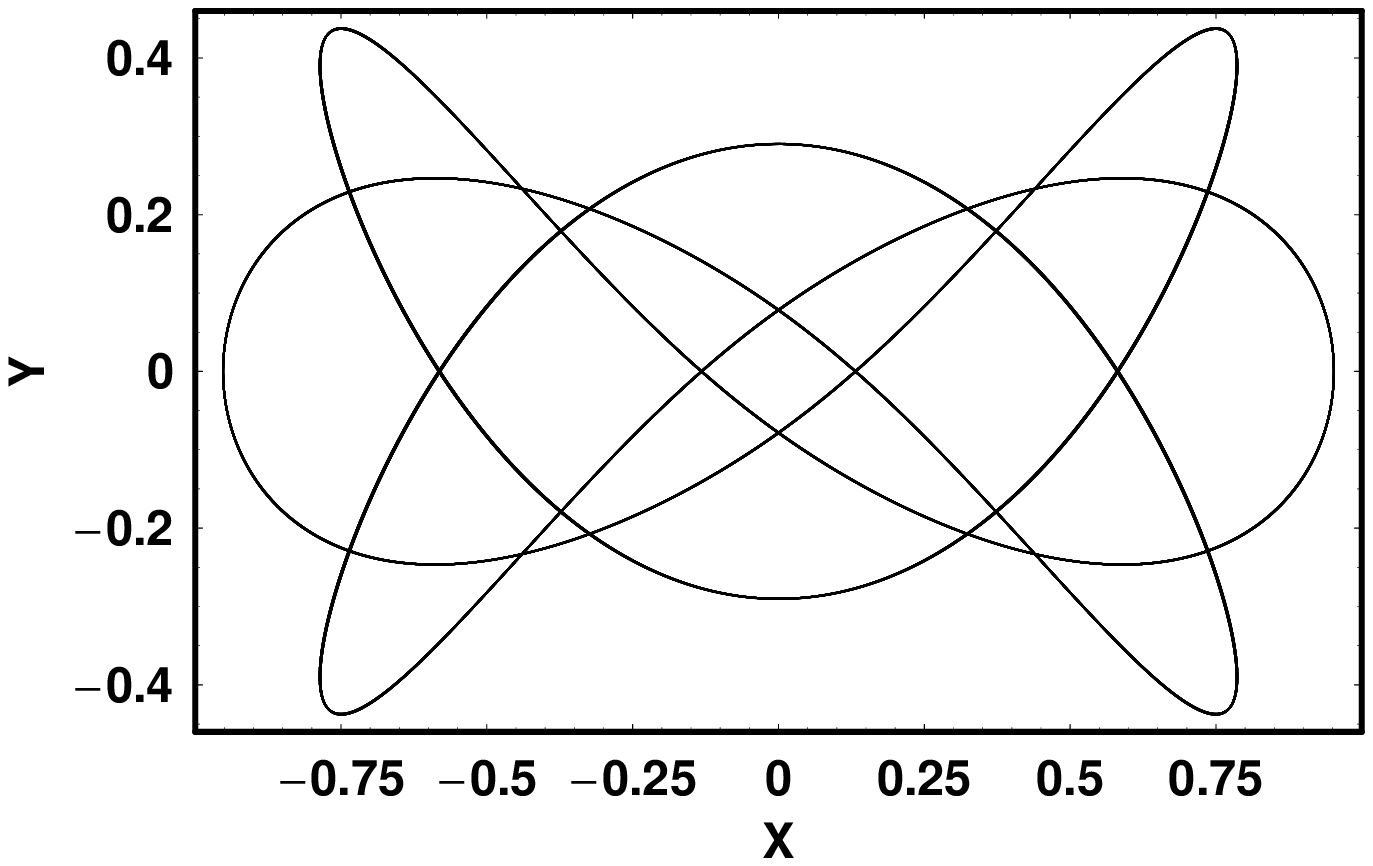}}}

\begin{minipage}{40mm}\centering

\hs {\fns(e)}
\end{minipage}\hspace{30mm}
\begin{minipage}{21mm}
\centering {\fns(f)}~~~~~~~~\end{minipage}

\resizebox{0.8\hsize}{!}{\rotatebox{0}{\includegraphics*{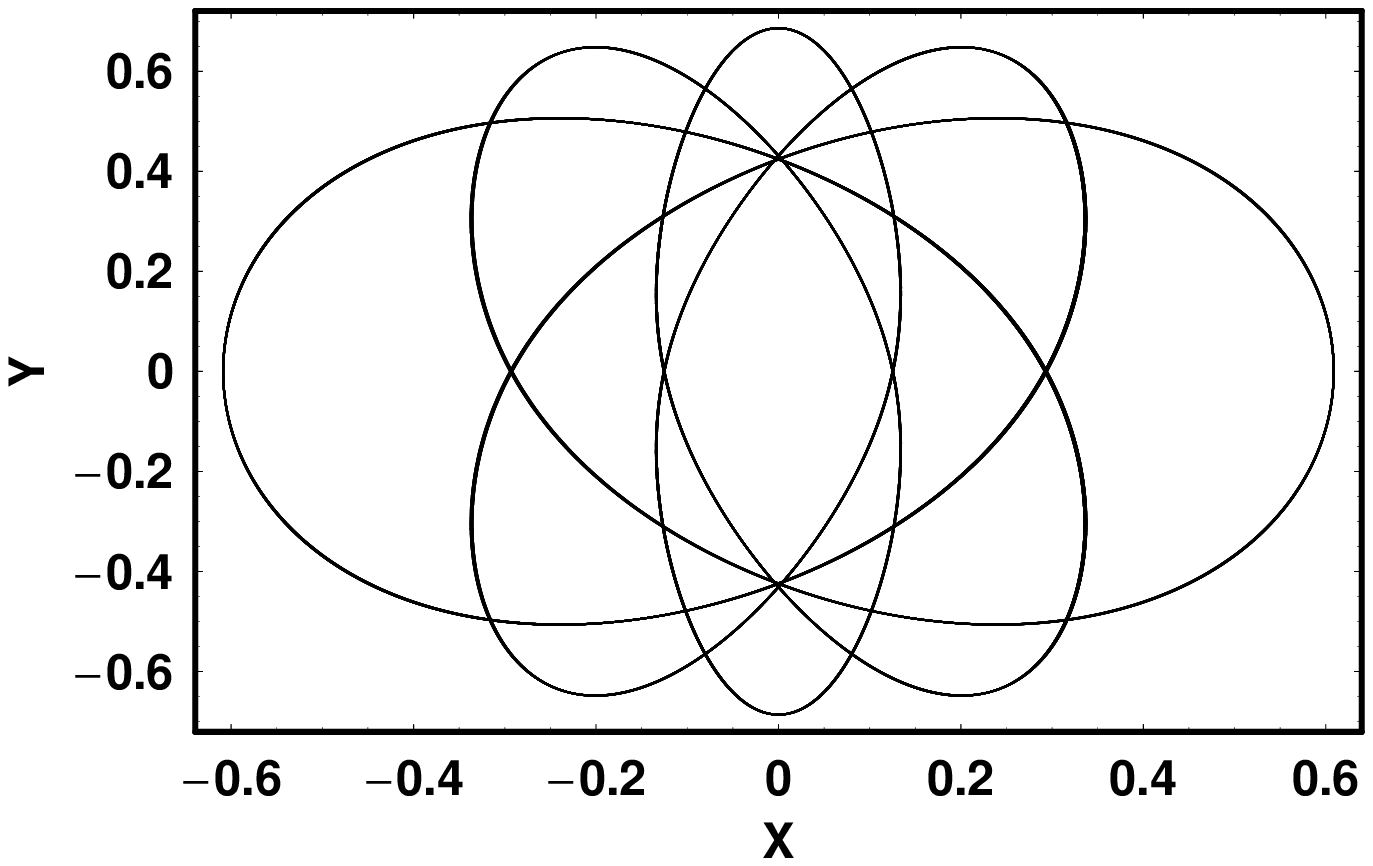}}
                         \rotatebox{0}{\includegraphics*{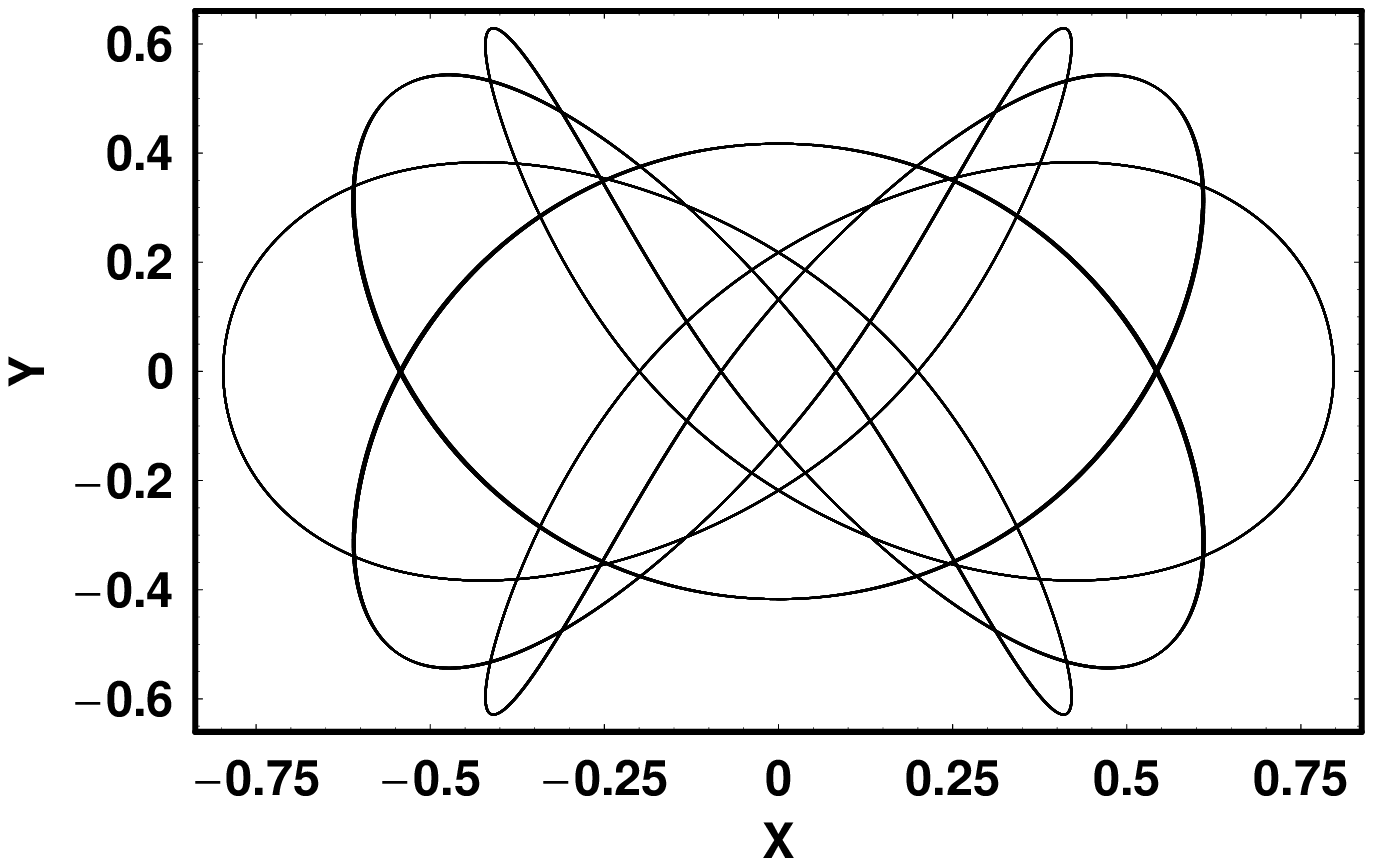}}}

\begin{minipage}{40mm}\centering

\hs {\fns(g)}
\end{minipage}\hspace{30mm}
\begin{minipage}{21mm}
\centering {\fns(h)}~~~~~~~~\end{minipage}

\resizebox{0.8\hsize}{!}{\rotatebox{0}{\includegraphics*{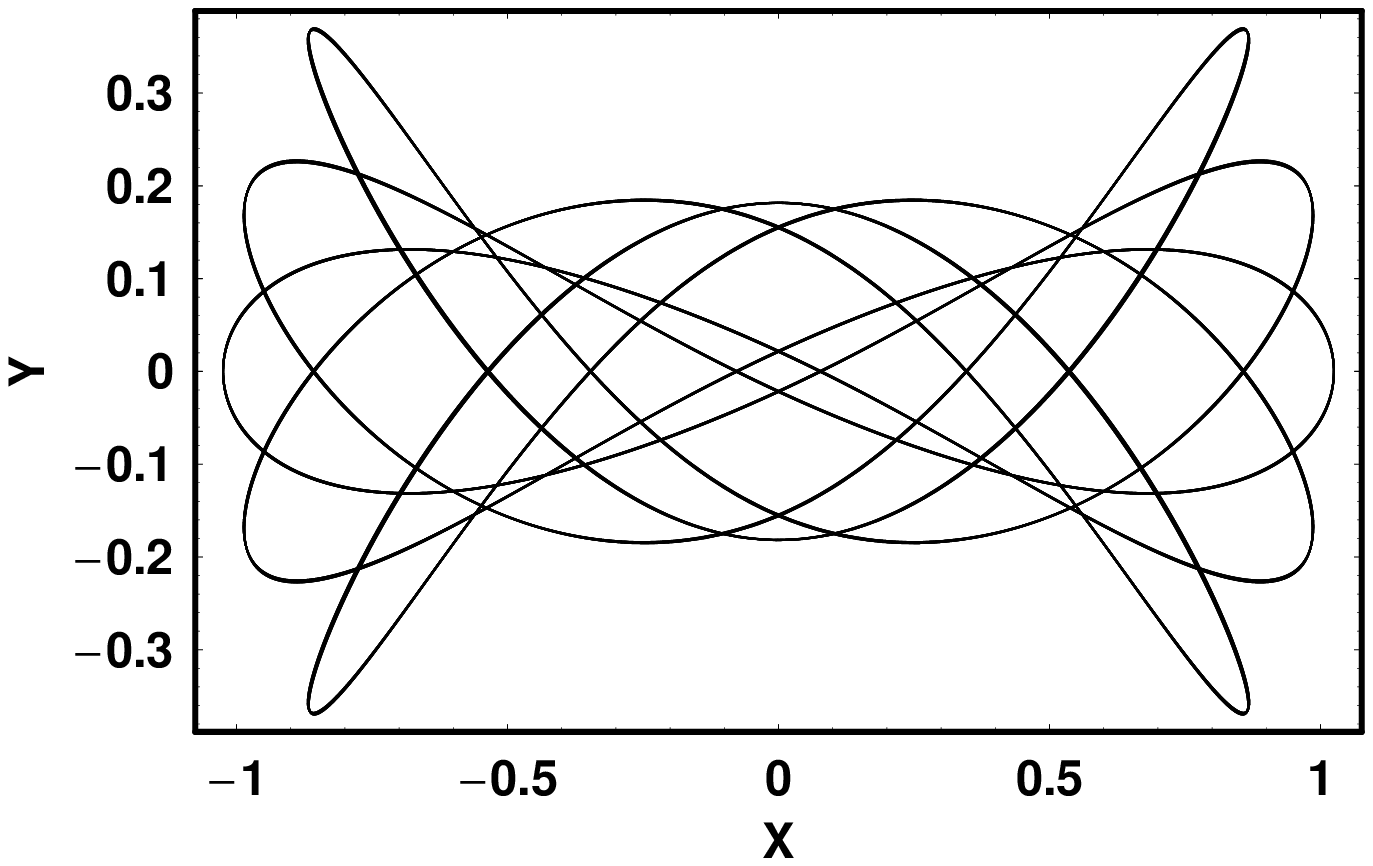}}
                         \rotatebox{0}{\includegraphics*{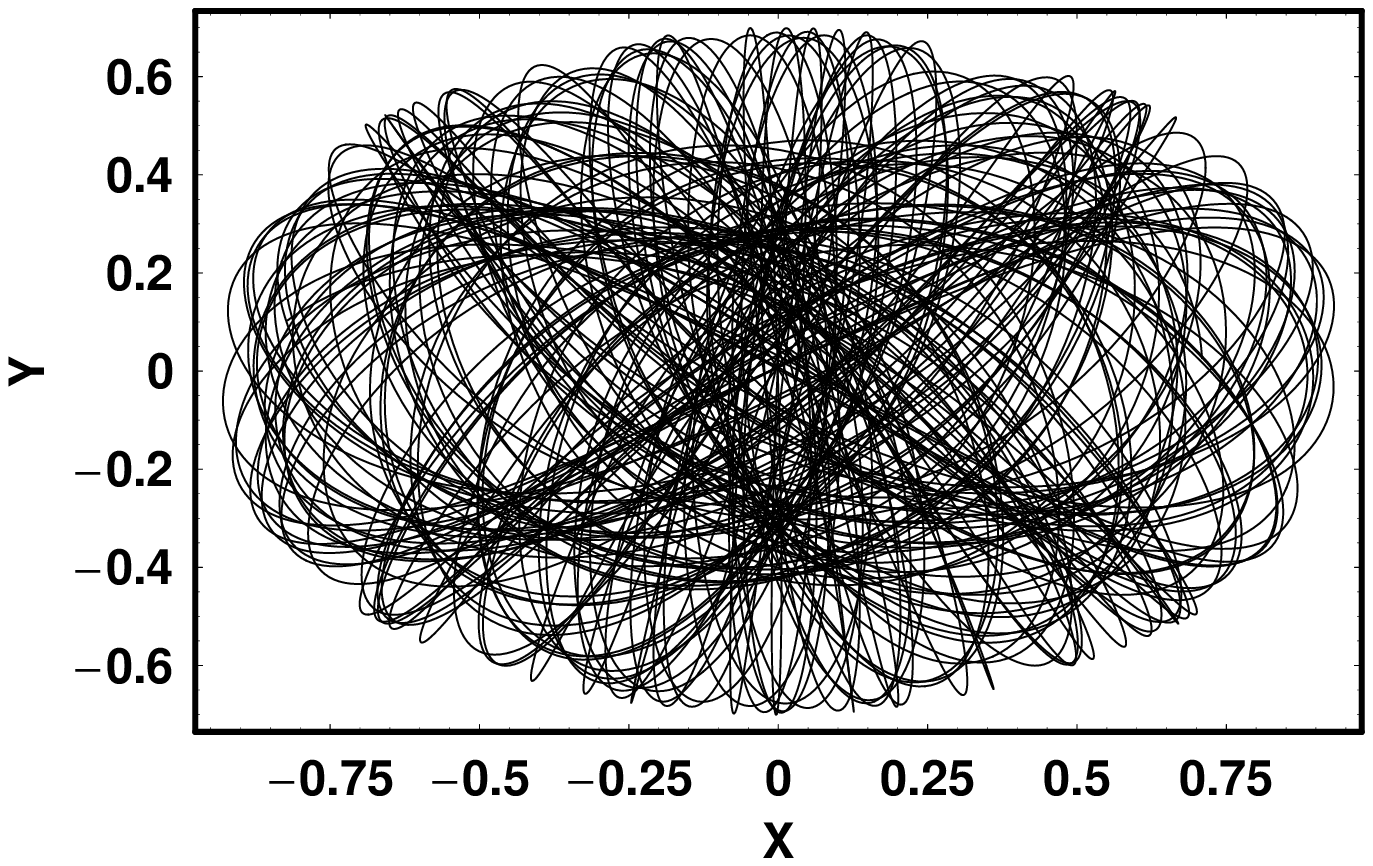}}}

\begin{minipage}{40mm}\centering

\hs {\fns(i)}
\end{minipage}\hspace{30mm}
\begin{minipage}{21mm}
\centering {\fns(j)}~~~~~~~~\end{minipage}
\begin{minipage}{115mm} \caption{\label{fig5}(a)--(j): Ten typical orbits in the
global Hamiltonian~(\ref{eq2}) when $E_{\rm J}=-2700$.}\end{minipage}
\end{figure*}

Let us now proceed in order to study the behavior of the orbits, in
the global model, near the galactic center, that is close to the
nuclear region. In order to visualize the nature of motion near the
nuclear region of our dynamical system, we present Figure~\ref{fig4}
which shows the $x-p_x$, $y=0$, $p_y>0$ phase plane when $E_{\rm
J}=-2700$. Once more, the particular value of the energy $E_{\rm J}$
was chosen so that in the phase plane $x_{\rm max} \simeq 1$. One
observes a very interesting and complicat{ed} phase
plane with regions of regular motion and a large, unified chaotic
domain. In fact, there are two main regular regions consisting of
invariant curves that are topological circles {en}closing the two
stable invariant points. The two stable invariant points $D$ and $R$
represent the direct (i.e. in the same direction as the rotation)
and the retrograde periodic orbits respectively. These periodic
orbits are characteristic of the 1:1 resonance and are similar to
ellipses circulating around the origin. The rest of the regular
region consists of several sets of smaller islands of invariant
curves produced by quasi-periodic orbits belonging to resonances of
higher multiplicity. The chaotic domain shown in the phase plane of
Figure~\ref{fig4} consists of a large, unified chaotic sea
and sticky regions. It is well known that the phenomenon of
stickiness is common in barred galaxies (see
\citealt{Caranicolas+Karanis+1998};
\citealt{Karanis+Caranicolas+2002}). The outermost curve shown in
the phase plane of Figure~\ref{fig4} is the limiting curve
defined by $H(x,p_x)=E_{\rm J}$.

Comparing the present results with those given in Paper P1, we
understand that the presence of the massive nucleus {dramatically
}changes the nature of motion in our barred galactic model. In Paper
P1, the motion was regular and there was only one kind of invariant
curve. The invariant curves were topological circles closing around
the unique central invariant point. The corresponding orbits were
box orbits. The outer invariant curves belong{ed} to elongated box
orbits that support the bar, {but} as we approach{ed} the central
invariant point, the box orbits bec{a}me more rectangular (see
fig.~4 of Paper P1 for more details). On the other hand, in the
present case, we have resonant periodic or quasi-periodic orbits
producing several sets of islands of invariant curves. Furthermore,
the majority of the phase plane shown in Figure~\ref{fig4} is
covered by chaotic orbits. We will return to this point later in the
discussion.

Figure~\ref{fig5}(a)--(j) shows ten typical orbits of the
Hamiltonian system, when $E_{\rm J}=-2700$. Figure~\ref{fig5}(a)
shows a periodic orbit starting at the position of the direct
stable invariant point. This orbit produces an invariant curve
which belongs to the main set of invariant curves and {encloses}
the fixed direct invariant point. The initial conditions are:
$x_0=-0.363$, $y_0=0$, $p_{x0}=0$. In Figure~\ref{fig5}(b) we see
a typical example of a 1:2 resonant periodic orbit, which produces
the set of the invariant curves close to the limiting curve. This
orbit has initial conditions: $x_0=0.51$, $y_0=0$, $p_{x0}=32.8$.
The orbit shown in Figure~\ref{fig5}(c) produces the set of two
small islands of invariant curves embedded near the main set of
invariant curves around the retrograde stable point. This orbit
represents the 2:2 resonance and has initial conditions:
$x_0=0.152$, $y_0=0$, $p_{x0}=0$. The orbit depicted in
Figure~\ref{fig5}(d) produces one of the islands of invariant
curves that intersects the $p_x$ axis. The initial conditions are:
$x_0=0.018$, $y_0=0$, $p_{x0}=38.6$. In Figure~\ref{fig5}(e) we
see a 3:4 resonant orbit with initial conditions: $x_0=0.0756$,
$y_0=0$, $p_{x0}=24.5$. Figure~\ref{fig5}(f) shows a periodic
orbit characteristic of the 3:5 resonance and produces a set of
five small islands of invariant curves embedded in the chaotic
region. This orbit has initial conditions: $x_0=0.952$, $y_0=0$,
$p_{x0}=0$. In Figure~\ref{fig5}(g) one can observe a typical 5:5
resonant periodic orbit. This orbit produces the set of the five
small islands around the invariant curves of the direct point and
has initial conditions: $x_0=-0.609$, $y_0=0$, $p_{x0}=0$.
Figure~\ref{fig5}(h) shows a periodic orbit which belongs to the
5:7 resonance family and produces a set of seven small islands of
invariant curves. The initial conditions are: $x_0=0.797, y_0=0,
p_{x0}=0$. In Figure~\ref{fig5}(i) we see a complicated periodic
orbit characteristic of the 5:9 resonance, which produces a set of
nine tiny islands embedded in the chaotic sea. The initial
conditions of this orbit are: $x_0=1.025, y_0=0, p_{x0}=0$.
Finally, in Figure~\ref{fig5}(j) we see a chaotic orbit with
initial conditions: $x_0=0.71, y_0=0, p_{x0}=0$. The initial value
of $p_{y0}$ was found in every case from the Jacobi
integral~(\ref{eq2}). All orbits shown in
Figure~\ref{fig5}(a)--(j) were calculated for a time period of 100
time units.

We have to point out that the nature of the local motion near
the central region of the galaxy is much more complicated than the
global motion. This can be justified by the fact that the phase
plane shown in Figure~\ref{fig4} {which} corresponds to
the local motion contains more resonant cases than the phase plane
of the global motion, which is shown in Figure~\ref{fig2}.
Therefore, we may conclude that the presence of a massive nucleus
near the galactic center gives rise to a large variety of
resonant orbits of higher multiplicity.

\section{Properties of motion in the local model}
\label{sect:prop}

The local potential can be found by expanding the effective
potential~(\ref{eq3}) in a Taylor series near the central stable
Lagrange point $L_1$, which coincides with the origin. Following
this procedure and keeping only terms up to the fourth degree in the
variables, we obtain the local effective potential which is
\begin{eqnarray}
U_{\rm eff}\left(\Delta x, \Delta y\right)
&=& U_{\rm eff}(0,0) + \frac{1}{2}
\left[A\left(\Delta x \right)^2 + B\left(\Delta y \right)^2\right] \nonumber \\
&&- \frac{1}{4}\left[\alpha_1\left(\Delta x \right)^4 + 2
\alpha_2\left(\Delta x \right)^2 \left(\Delta y \right)^2 +
\alpha_3\left(\Delta y \right)^4\right] \nonumber\\
 &&-
\frac{1}{2}\Omega_0^2 \left[\left(\Delta x \right)^2 +
\left(\Delta y \right)^2\right], \label{eq8}
\end{eqnarray}
where we have set
\begin{equation}
U_{\rm eff} = \frac{\alpha c_{\rm n}^2}{M_{\rm d}} \Phi_{\rm eff},
\end{equation}
in order to avoid large numbers. Setting for convenience: $x=\Delta x$, $y=\Delta y$ and $V_{\rm eff}=U_{\rm eff}-U_{\rm eff}(0,0)$, Equation~(\ref{eq8}) becomes
\begin{equation}
V_{\rm eff}(x,y) = \frac{1}{2} \left(A x^2 + B y^2 \right) - \frac{1}{4} \left(\alpha_1 x^4 + 2\alpha_2 x^2 y^2 + \alpha_3 y^4 \right) - \frac{1}{2} \Omega_0^2 \left(x^2 + y^2 \right),
\label{eq10}
\end{equation}
where
\begin{eqnarray}
A &=& \frac{c_{\rm n}^2}{\alpha^2} + \frac{\alpha
c_{\rm n}^2 M_{\rm b}}{M_{\rm d} c_{\rm b}^3} +
\frac{\alpha M_{\rm n}}{M_{\rm d} c_{\rm n}} +
\frac{\upsilon_0^2 \alpha c_{\rm n}^2}{M_{\rm d} c_{\rm h}^2}\, , \nonumber \\
B &=& \frac{c_{\rm n}^2}{\alpha^2} + \frac{\alpha b^2 c_{\rm n}^2 M_{\rm b}}{M_{\rm d} c_{\rm b}^3} + \frac{\alpha M_{\rm n}}{M_{\rm d} c_{\rm n}} + \frac{\upsilon_0^2 \alpha \beta c_{\rm n}^2}{M_{\rm d} c_{\rm h}^2}, \nonumber \\
\alpha_1 &=& \frac{3}{2}\left(\frac{c_{\rm n}^2}{\alpha^4} +
 \frac{\alpha M_{\rm b}}{M_{\rm d} c_{\rm b}^5} +
 \frac{\alpha M_{\rm n}}{M_{\rm d} c_{\rm n}^3}\right) +
  \frac{\upsilon_0^2 \alpha c_{\rm n}^2}{M_{\rm d} c_{\rm h}^4}\, , \nonumber \\
\alpha_2 &=& \frac{3}{2}\left(\frac{c_{\rm n}^2}{\alpha^4} + \frac{\alpha b^2 c_{\rm n}^2 M_{\rm b}}{M_{\rm d} c_{\rm b}^5} + \frac{\alpha M_{\rm n}}{M_{\rm d} c_{\rm n}^3}\right) + \frac{\upsilon_0^2 \alpha \beta c_{\rm n}^2}{M_{\rm d} c_{\rm h}^4}, \nonumber \\
\alpha_3 &=& \frac{3}{2}\left(\frac{c_{\rm n}^2}{\alpha^4} +
\frac{\alpha b^4 c_{\rm n}^2 M_{\rm b}}{M_{\rm d} c_{\rm b}^5} +
 \frac{\alpha M_{\rm n}}{M_{\rm d} c_{\rm n}^3}\right) +
  \frac{\upsilon_0^2 \alpha \beta^2 c_{\rm n}^2}{M_{\rm d} c_{\rm h}^4}\, , \nonumber \\
\Omega_0 &=& c_{\rm n} \Omega_{\rm b} \sqrt{\frac{\alpha}{M_{\rm
d}}}\, . \label{eq11}
\end{eqnarray}
As one can se{e} from Equation~(\ref{eq11}), the coefficients of
the local effective potential are functions of the physical
quantities entering the global effective potential~(\ref{eq3}).

The local Hamiltonian is
\begin{equation}
H_{\rm L} = \frac{1}{2}\left(p_x^2 + p_y^2\right) + V_{\rm
eff}(x,y) = h_{\rm L}\, , \label{eq12}
\end{equation}
where $p_x$ and $p_y$ are the local momenta per unit mass,
conjugate to $x$ and $y$ respectively, while $h_{\rm L}$ is the
numerical value of the local energy.

In order to connect the value of the global energy $E_{\rm J}$ with
the value of the local energy $h_{\rm L}$, we proceed as follows.
The equation $E_{\rm J0}=\Phi_{\rm eff}(0,0)$ defines a point in the
$\left(x,y\right)$ plane, while $E_{\rm J}=\Phi_{\rm eff}(x,y)$
defines a curve in the same plane. The global motion takes place
inside this curve, which is known as the zero velocity curve. At the
same time, $h_{\rm L0}=V_{\rm eff}(0,0)$ defines a point in the
$\left(x,y\right)$ plane, while $h_{\rm L}=U_{\rm eff}(x,y)$ defines
a curve inside which the local motion takes place. This second curve
is the local zero velocity curve. We {only }consider bounded motion,
{which means} the zero velocity curves are always closed. The local
energy $h_{\rm L}$ is connected to the global energy $E_{\rm J}$
through the relation
\begin{eqnarray}
h_{\rm L} &=& U_{\rm eff}(x,y) - U_{\rm eff}(0,0) = \frac{\alpha c_{\rm n}^2}{M_{\rm d}} \left[\Phi_{\rm eff}(x,y)
- \Phi_{\rm eff}(0,0)\right] \nonumber \\
&=& \frac{\alpha c_{\rm n}^2}{M_{\rm d}} \left(E_{\rm J} - E_{\rm J0} \right),
\label{eq13}
\end{eqnarray}
where
\begin{equation}
E_{\rm J0} = - \left(\frac{M_{\rm d}}{\alpha} + \frac{M_{\rm b}}{c_{\rm b}} + \frac{M_{\rm n}}{c_{\rm n}} \right)
+ \upsilon_0^2 \ln\left(c_{\rm h}\right).
\end{equation}

Let us now study the properties of the local motion. For the adopted
values of the global parameters and for $\Omega_{\rm b}=1.25$, we
find that: $A = 1.4, B = 1.5, \alpha_1 = 32.8, \alpha_2 = 32.4,
\alpha_3 = 32.8$ and $\Omega_0 = 0.009$. For the value of the global
energy $E_{\rm J}=-2700$ we find, by using Equation~(\ref{eq13}),
that $h_{\rm L} = 0.08452$. It is amazing that this value of the
local energy is much larger than the energy of escape for
the corresponding local potential (see
\citealt{Caranicolas+Varvoglis+1984};
\citealt{Caranicolas+Karanis+1998}), which is given by
\begin{equation}
h_{\rm esc} = \frac{\left(B - \Omega_0^2\right)^2}{6\alpha_1}\, .
\label{eq15}
\end{equation}

For the above values of the parameters, Equation~(\ref{eq15})
gives the value $h_{\rm esc} = 0.01196$. This value is less than
$h_{\rm L}/7$. Therefore, one may conclude that near the massive
nucleus we do not have local motion, or equivalent{ly} local
orbits escape the nuclear region, because they possess a high
value of local energy. In order to obtain an idea regarding the
nature of the local motion near the nucleus, we must go to very
low values of energies, that is when $h \ll h_{\rm L}$, where $h$
is the value of the energy.
\begin{figure}
\centering
\resizebox{0.6\hsize}{!}{\rotatebox{270}{\includegraphics*{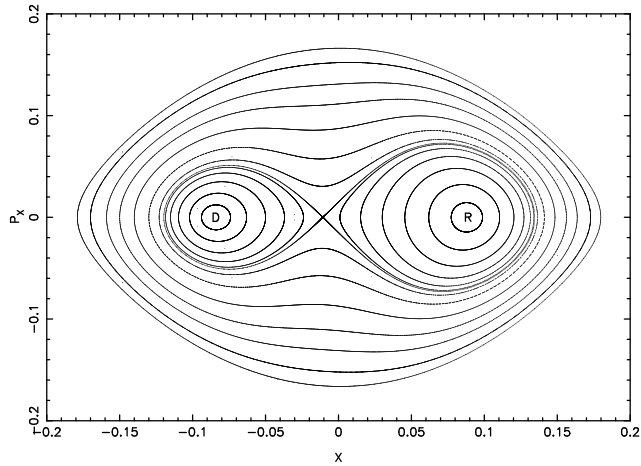}}}
\caption{\label{fig6}The $x-p_x$ Poincar\'{e} phase plane for the
local Hamiltonian~(\ref{eq12}), when $h=0.012$.}
\end{figure}

\begin{figure*}
\centering
\resizebox{0.8\hsize}{!}{\rotatebox{0}{\includegraphics*{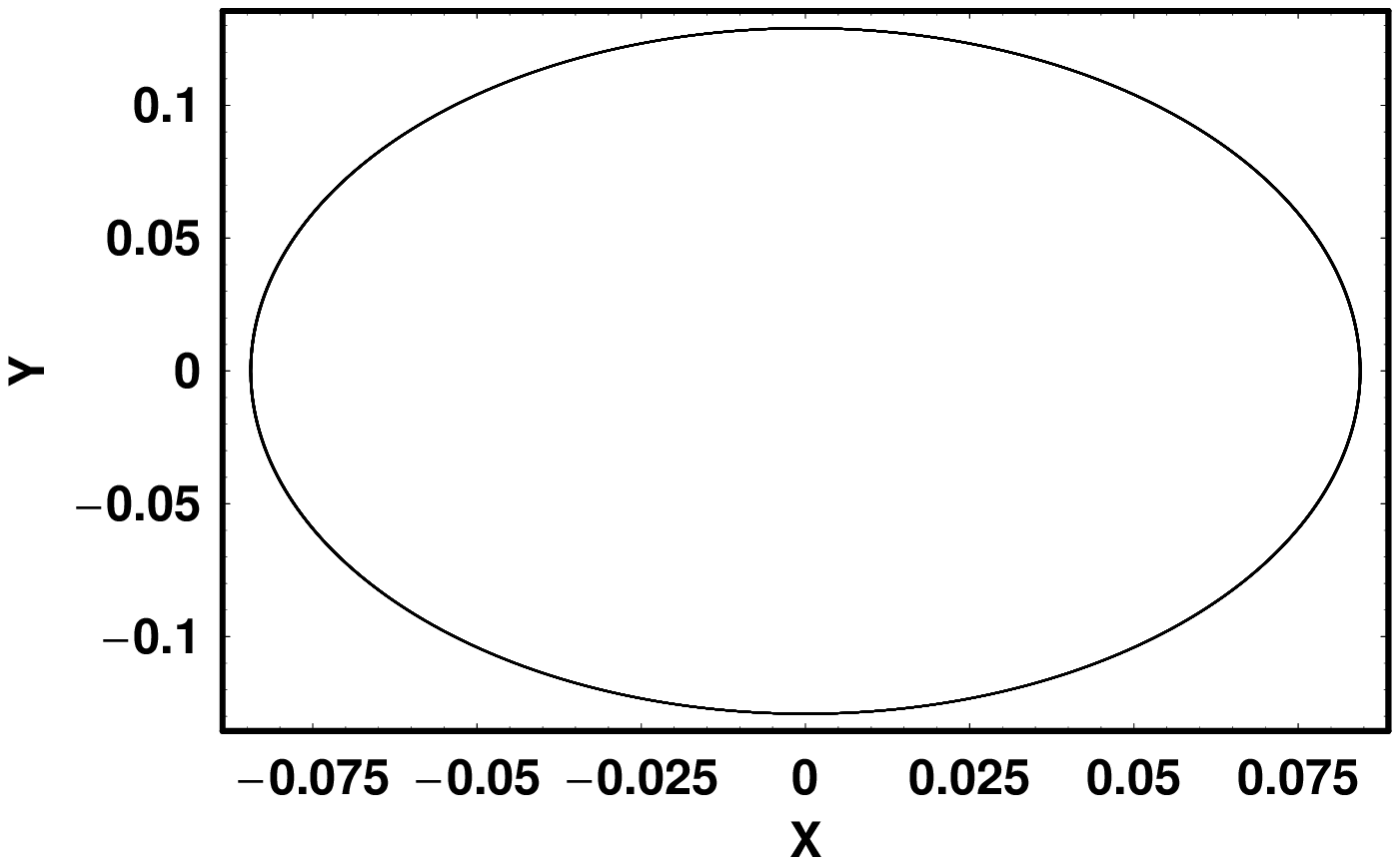}}
                         \rotatebox{0}{\includegraphics*{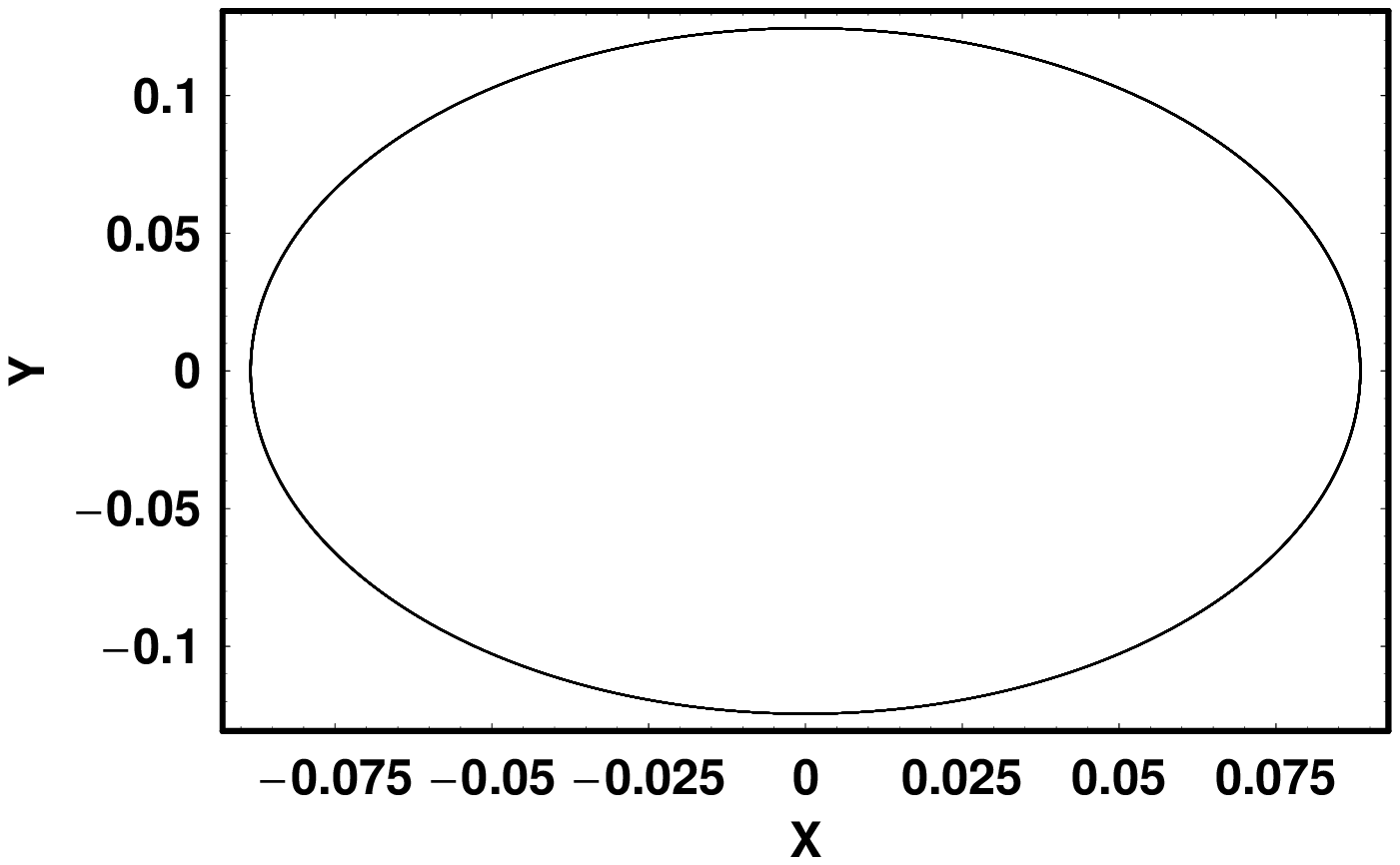}}}

\begin{minipage}{40mm}\centering

\hs {\fns(a)}
\end{minipage}\hspace{30mm}
\begin{minipage}{21mm}
\centering {\fns(b)}~~~~~~~~\end{minipage}

\resizebox{0.8\hsize}{!}{\rotatebox{0}{\includegraphics*{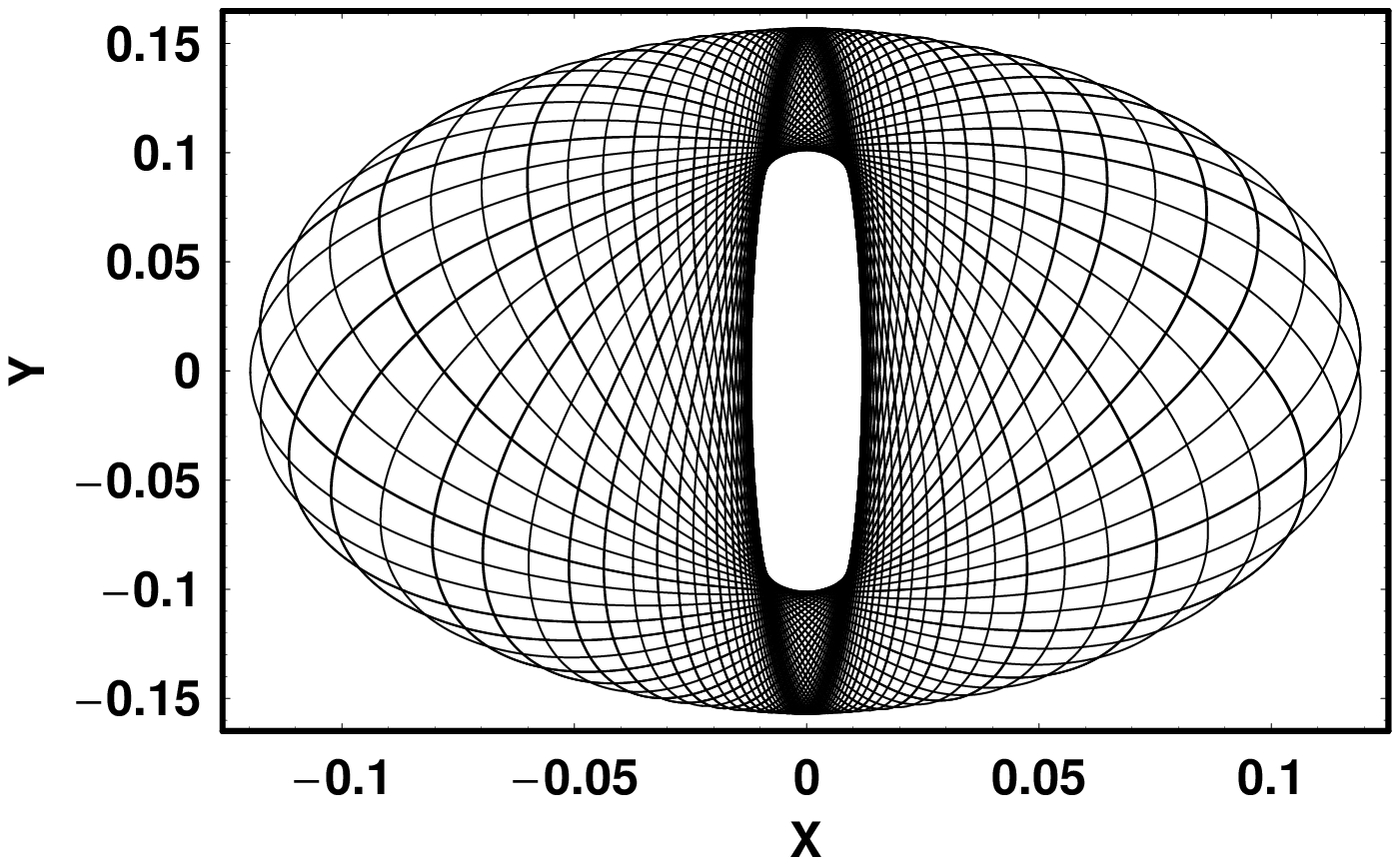}}
                         \rotatebox{0}{\includegraphics*{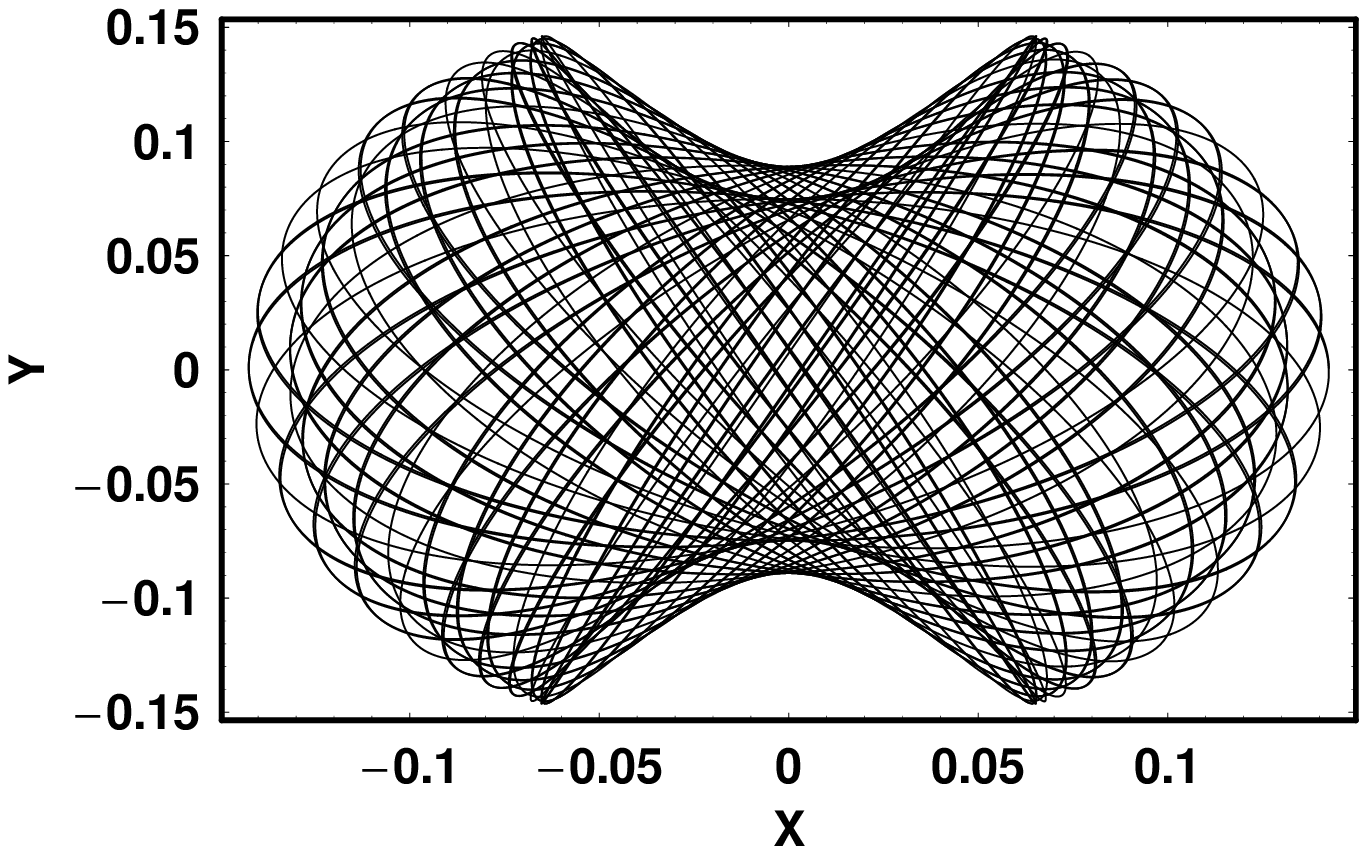}}}

\begin{minipage}{40mm}\centering

\hs {\fns(c)}
\end{minipage}\hspace{30mm}
\begin{minipage}{21mm}
\centering {\fns(d)}~~~~~~~~\end{minipage} \vspace{-3mm}
\begin{minipage}{115mm}\caption{\label{fig7}(a)--(d): Four typical
orbits in the local Hamiltonian~(\ref{eq12}), when
$h=0.012$.}\end{minipage}
\end{figure*}

Figure~\ref{fig6} shows the $x-p_x$, $y=0$, $p_y>0$ Poincar\'{e}
phase plane when $h = 0.012$. The motion is regular and the phase
plane has all the characteristics of the 1:1 resonance. There are
two stable invariant points marked as $D$ and $R$ corresponding to
direct and retrograde resonant periodic orbits respectively. The
unstable periodic point gives an orbit, which in the absence of
rotation is the $p_x$ axis. Thus, we observe that the local motion
near the nucleus consists of a low energy 1:1 resonant periodic
orbit. Moreover, as the entire phase plane is covered with invariant
curves corresponding only to regular orbits{,} we conclude that near
the nucleus we do not observe any kind of local chaotic motion. This
situation is completely different from that displayed in Paper P1,
where the nucleus was not present. In that case, all orbits in the
local potential were box orbits, without any resonance phenomena.

Figure~\ref{fig7}(a)--(d) shows four orbits of the local
Hamiltonian~(\ref{eq12}). Figure~\ref{fig7}(a) and (b) show two
periodic orbits starting at the direct and the retrograde periodic
points respectively. The initial conditions for the orbit shown in
Figure~\ref{fig7}(a) are: $x_0=-0.0845$, $y_0=0$, $p_{x0}=0$, while
for the orbit shown in Figure~\ref{fig7}(b) are: $x_0=0.0885$,
$y_0=0$, $p_{x0}=0$. In Figure~\ref{fig7}(c) we observe an orbit
starting at the unstable periodic point with initial conditions:
$x_0=-0.0119$, $y_0=0$, $p_{x0}=0$. Finally, in Figure~\ref{fig7}(d)
we see a box orbit which produces one of the outer invariant curves
shown in the phase plane of Figure~\ref{fig6}. The initial
conditions for this orbit are: $x_0=0.1425$, $y_0=0$, $p_{x0}=0$.
The initial value of $p_{y0}$ was found in every case from the
energy integral~(\ref{eq12}). All orbits shown in
Figure~\ref{fig7}(a)--(d) were calculated for a time period of 100
time units.

We must clarify to the reader that the Taylor
expansion~(\ref{eq10}) is only valid when
\begin{equation}
\frac{x^2 + y^2}{\alpha^2} \ll 1, \qquad \frac{x^2 + b^2
y^2}{c_{\rm b}^2} \ll 1, \qquad \frac{x^2 + y^2}{c_{\rm n}^2} \ll
1, \qquad \frac{x^2 + \beta y^2}{c_{\rm h}^2} \ll 1. \label{eq16}
\end{equation}
It is also important to note that, for a given value of the global
energy $E_{\rm J}$, a corresponding value of the local energy
$h_{\rm L}$ can be obtained through relation~(\ref{eq13}). It is
obvious that this value of the local energy does not have any
physical meaning if all the relations~(\ref{eq16}) are not
satisfied.

\begin{figure}[!tH]
\centering
\resizebox{0.6\hsize}{!}{\rotatebox{0}{\includegraphics*{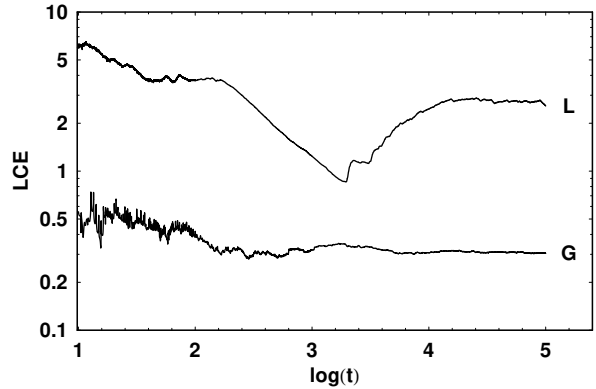}}}
\caption{\label{fig8}\baselineskip 3.6mm Time evolution of the
L.C.E{.} for two chaotic orbits. The curve labeled $G$
corresponds to an orbit in the chaotic domain of the phase plane of
Fig.~\ref{fig2}, while the curve labeled $L$ corresponds to an orbit
in the chaotic domain of the phase plane of Fig.~\ref{fig4}.}
\end{figure}

In order to {better }estimate the degree of chaos displayed by the
chaotic orbits shown in Figures~\ref{fig2} and \ref{fig4}, we
decided to compute the maximum Lyapunov Characteristic Exponent
(L.C.E{.}) (for details see
\citealt{Lieberman+Lichtenberg+1992}). The results are shown in
Figure~\ref{fig8}. The curve labeled $G$ shows the evolution of the
L.C.E{.} for a chaotic orbit with initial conditions in the
chaotic region of Figure~\ref{fig2}. The particular values of the
initial conditions of this orbit are as in
Figure~\ref{fig3}(f). On the other hand, the curve labeled $L$
shows the evolution of the L.C.E{.} for a chaotic orbit with
initial conditions in the chaotic region of Figure~\ref{fig4}. The
particular values of the initial conditions of this orbit are
as in Figure~\ref{fig5}(j). We observe that the value of the
L.C.E{.} corresponding to the local system of Figure~\ref{fig4}
is about ten times the value of the L.C.E{.} corresponding to
the global system of Figure~\ref{fig2}. Therefore, one can say that
in this case we have not only fast chaos (see
\citealt{Caranicolas+Vozikis+1987}), where the L.C.E{.} was {on the}
order of unity, but{ also} very fast chaos where the L.C.E{.} is about
three times larger. It is evident that this is an indication
of strong nuclear activity near the vicinity of the galactic
center.

\section{Discussion and conclusions}
\label{sect:discus}

One of the most important approaches in order to understand and
reveal the dynamical behavior of a galactic system is based on the
knowledge of the chaotic versus ordered nature of orbits. In the
present research, we tried to achieve this using a global and a
local potential describing a barred galaxy with a massive nucleus, a
disk and prolate halo components. The corresponding local potential
was found by expanding the global potential around the central
stable Lagrange point, which coincides with the origin, in a Taylor
series and keeping only terms up to the fourth degree in the
variables. This local potential is a potential made up of a
two-dimensional perturbed harmonic oscillator. The study of the
motion in those potentials has been an active field of research over
the last decades (see for examples \citealt{Saito+Ichimura+1979};
\citealt{Innanen+1985}; \citealt{Caranicolas+1984, Caranicolas+1994,
Caranicolas+2000}; \citealt{Caranicolas+Karanis+1999};
\citealt{Caranicolas+Vozikis+2002}). In {later} years, the study of
the properties of motion in those dynamical systems has been
explored in detail, using precise and modern analytical (see
\citealt{Elipe+1999, Elipe+2001}; \citealt{Elipe+Deprit+1999}) or
numerical (see \citealt{Lara+etal+1999};
\citealt{Karanis+Caranicolas+2002}) methods.

As expected, the local parameters and the corresponding local energy
are functions of all the {involved }global parameters and the global
energy. Our numerical experiments, in the global dynamical system
for high values of the energy, suggest that more than $62\%$ of the
tested orbits are chaotic. Comparing the outcomes with those of
Paper P1 we see that, in the present case, we have a sharp increase
of the total chaotic motion. It is evident that the chaotic domain
in Figure~\ref{fig2} is much more extended. This means that the
additional nucleus affects not only the region near the center of
the galaxy, but also areas far from {it}. Another interesting
observation is that the presence of the nucleus does not seem to
affect the figure-eight quasi-periodic orbits, which are the
building blocks of the barred structure of the galaxy. This result
agrees with recent observations made by the Hubble Space
Telescope, which revealed that a number of Seyfert galaxies
display strong nuclear barred structure (see
\citealt{Regan+Mulchaey+1999}). On the other hand, the global motion
near the massive nucleus has the characteristic of {a} 1:1
resonance. Furthermore, the chaotic domain seems to be larger than
that observed in Figure~\ref{fig2}. This is natural because of the
presence of the nearby massive nucleus. All the above results are
completely different from the corresponding results obtained in
Paper P1, where the motion was regular and all the orbits in the
global system near the galactic center were box orbits.

The effect of a central mass concentration (CMC) on the phase
space of global galactic models has already been studied
extensively in several earlier research works (see Caranicolas \&
Zotos 2011; Zotos 2012).
 \cite{Hasan+Norman+1990} have shown that a black hole or a CMC
in a barred galaxy can dissolve the barred structure. Moreover,
the fundamental orbits of the system (called B orbits), supporting
the bar{,} are orbits elongated in the direction of the bar. As
the black hole's mass increases, an inner Lindblad resonance (ILR)
appears and moves outward. Thus, the B orbits disappear when the
ILR reaches the end of the bar. This effect is significantly
increased if the bar is thinner. Orbits that stay closer to the
central mass, with smaller Jacobi constants, can change their
character from regular to chaotic more easily. Furthermore, the
study of \cite{Hasan+etal+1993} indicates that, as the mass of the
CMC increases{,} the stable regular orbits in the region where the
bar potential competes in strength with the central mass
potential, especially around the region of the ILR{,} will no
longer {be }present and their position will be occupied by chaotic
orbits. \cite{Shen+Sellwood+2004} have conducted a systematic
study of the effects of CMC on bars, using high quality $N$-body
simulations. They have experimented with both strong and weak
initial bars and a wide range of the physical parameters of the
CMC, such as the final mass, the scale length and the mass growth
time. Their outcomes suggest that, for a given mass{,} compact
CMCs (such as super-massive black holes) are more destructive to
barred structures than are more diffuse ones (such as molecular
gas clouds in many galactic centers). They have shown that the
former are more efficient scatter{er}s of bars supporting regular
elongated orbits, that pass close to the galactic center and{,}
therefore, decrease the percentage of the regular orbits and
increase the area of the chaotic region in the phase space. All
the above research outcomes are in agreement with the present
findings, which strongly indicate that the massive nucleus is a
very important parameter of the dynamical system, since {it }is
responsible not only for the chaotic motion of low energy stars
but also for the several resonant orbits of higher multiplicity
that appear in the local model. In particular,
Figure{s}~\ref{fig2} and \ref{fig4} in the current paper
remarkably{ resemble} the Poincar\'{e} surface of section plots in
figure~11 {of} \cite{Shen+Sellwood+2004}, where a realistic
self-consistent barred model was adopted.

Let us now consider the local motion. The local
potential~(\ref{eq10}) has all the characteristics of the 1:1
resonance. Such potentials are known as perturbed elliptic
oscillators (see \citealt{Deprit+1991}). It is very interesting that
for the corresponding local energy given by Equation~(\ref{eq13}),
the local motion {is} not bounded. In order to obtain
bounded motion, one must approach very low values of local energies.
Strictly speaking, the presence of the massive nucleus makes the
local energy increase dramatically, so that considering all
the zero velocity curves results in unbounded local motion.
For values of local energies slightly lower than the value of the
energy of escape, we observe resonance phenomena, namely the 1:1
resonant periodic orbits. Extensive numerical calculations in the
local potential suggest that chaotic motion was not observed.

Bearing all the above in mind, we can say that this situation is
entirely different from that investigated in Paper P1, where the
properties of local motion were the same as those of the global
motion near the galactic center (see figs. 4 and 5 in Paper P1). In
our case, the corresponding local motion does not appear to exist
and only very low energy local motion seems to be present.
Furthermore, the fact that the corresponding value of the
L.C.E{.}, which was computed in the chaotic domain near the
nuclear region, is much larger than unity suggests
that there is a particular and strong activity in the central parts
of galaxies with strong nuclear bars.

We consider this paper to be an initial effort in order to explore
and reveal the dynamical structure of the system in more detail. As
the present results are positive, further investigation will be
initiated to study all the available phase space. Moreover, our
gravitational model will be suitably modified, in order to be able
to describe the properties of motion in a Hamiltonian galactic
system of three 
degrees of freedom.

\begin{acknowledgements}
{The author would} like to express his thanks to the anonymous
referee for the careful reading of the manuscript and for his very
useful suggestions and comments, which improved the quality of the
present work.
\end{acknowledgements}

\end{document}